# Chiral Nonlinear Polaritonics with van der Waals Metasurfaces

Connor Heimig*[1], Alexander A. Antonov*[1], Dmytro Gryb[1], Thomas Possmayer[1], Thomas Weber[1], Michael Hirler[1], Jonas Biechteler[1], Luca Sortino[1], Leonardo de S. Menezes[1,2], Stefan A. Maier[3,4], Maxim V. Gorkunov[5,6], Yuri Kivshar[7], and Andreas Tittl†[1]

[1] *Chair in Hybrid Nanosystems, Nanoinstitute Munich, Faculty of Physics, Ludwig-Maximilians-Universität München, Munich, Germany*
[2] *Departamento de Física, Universidade Federal de Pernambuco, Recife-PE, Brazil*
[3] *School of Physics and Astronomy, Monash University, Clayton, Victoria, Australia*
[4] *Department of Physics, Imperial College London, London, UK*
[5] *Shubnikov Institute of Crystallography, National Research Centre "Kurchatov Institute", Moscow, Russia*
[6] *Theoretical Physics and Quantum Technologies Department, National University of Science and Technology 'MISIS', Moscow, Russia*
[7] *Nonlinear Physics Centre, Research School of Physics, Australian National University, Canberra, Australia*

In the strong-coupling regime, the interaction between light and matter reaches a hybridization state where the photonic and material components are inseparably linked. Using tailored states of light to break symmetries in such systems can facilitate the development of novel non-equilibrium quantum materials. Chiral optical cavities offer a promising approach for this, enabling either temporal or spatial symmetry-breaking, both of which are unachievable with conventional mirror cavities. For spatial symmetry-breaking, a cavity must discriminate the handedness of circularly polarized light, a functionality uniquely provided by chiral metamaterials. Here, we propose and demonstrate experimentally a chiral transition metal dichalcogenide (TMDC) metasurface with broken out-of-plane symmetry, allowing for the selective formation of self-hybridized exciton-polaritons with specific handedness. Our metasurface maintains maximal chirality for oblique incidence up to 20°, significantly outperforming all previously known designs, thereby transforming the angle of incidence from a constraint into a new degree of freedom for sub-nanometer-precise tuning of the cavity's resonant wavelength. Moreover, we study the chiral strong-coupling regime in nonlinear experiments and reveal the polariton-driven nature of chiral third-harmonic generation. Our results demonstrate a clear pathway towards van der Waals (vdW) metasurfaces as a novel and potent platform for chiral polaritonics with implications in a wide range of photonics research, such as non-reciprocal photonic devices and valleytronics.

---

*These authors contributed equally to this work.
†Andreas.Tittl@physik.uni-muenchen.de

## Introduction

The interaction of chiral light and matter has become a significant area of interest within the broader nanophotonics community, including chiral sensing [1, 2], quantum emission [3], and photonic circuitry [4]. Recently, the emerging class of two-dimensional (2D) materials has become a focal point for these investigations, with a particular emphasis on the family of transition metal dichalcogenides (TMDCs).[5, 6] The role of chirality is especially prominent in TMDC research because it plays a crucial role in the formation of excitonic valleys in TMDC monolayers.[7] These atomically layered materials exhibit considerable application potential due to their electronic and optical properties, which have been utilized in a diverse range of research areas, including spintronics [8], electrocatalysis [9], topological insulators [10], and superconductivity [11].

A defining feature of these materials is their pronounced excitonic properties, particularly when examined in the context of strong coupling.[12] This denotes a condition wherein the interaction between light and matter becomes so intense that they can no longer be regarded as independent entities. Instead, they give rise to novel hybrid quasi-particles, known as polaritons, exhibiting attributes from both underlying components.[13] These hybrid properties bring about a host of intriguing phenomena, with a wide range of applications from quantum computing [14] to lasing [15]. To achieve strong coupling, a frequent approach is to place an excitonic system within or in proximity to an optical cavity, which provides the requisite confined light mode.[16] Chiral polaritonics has the potential to become a new frontier for nanophotonics [17], with first successful experimental realizations based on chiral plasmonic-excitonic systems [18] and DNA Origami [19]. Previous theoretical investigations provided initial insights into the chiral strong-coupling picture by combining material chirality with excitons.[20] The impact of a chiral cavity on graphene was also studied theoretically, displaying the quantized light-induced anomalous Hall-effect.[21] Still, the experimental development of more generally applicable chiral polaritonic platforms that move beyond material-intrinsic chirality remains a crucial and persistent need.

Combining conventional optical resonators, such as mirror-based Fabry–Pérot cavities, with excitonic materials enables strong coupling between photonic modes and excitons [22]. However, generating the chiral optical fields necessary for chiral strong coupling typically requires elaborate configurations, such as precisely aligned pairs of chiral mirrors.[23, 24] Here, we propose an alternative approach: a single metasurface that simultaneously supports a high-quality-factor chiral optical resonance and incorporates a TMDC material with strong excitonic response. Metasurfaces are artificially nano-engineered planar structures that have been specifically designed to manipulate light by periodically arraying sub-wavelength building blocks, enabling a wide range of optical functionalities.[25–27] When aiming to confine a light mode in a metasurface, the concept of bound states in the continuum (BIC) has emerged as a promising candidate.[28] This allows precise control over the radiative losses in the system as well as the confinement of ultra-sharp resonances. Such BIC-metasurfaces serve the same function in polaritonics as a cavity, providing the near-field enhancement necessary to strongly couple light with an exciton. In this way, they function analogously to open cavity systems but in a significantly simpler and more compact form, as strong coupling occurs within a much smaller volume and does not require opposing reflectors. Furthermore, metasurfaces have been shown to support maximally chiral eigenstates that are uncoupled from light of a particular circular polarization while resonantly interacting with its opposite-handed counterpart.[29, 30]

Previous work aimed at coupling excitons in TMDCs to a confined BIC mode was based on the transfer or growth of a single atomic layer onto a metasurface fabricated from conventional dielectric materials.[31] This approach suffers from numerous inherent limitations, from scalability to strain-driven alterations to the atomic layers' properties.[32] The principal benefit of our monolithic methodology (where the entire metasurface is fabricated from bulk TMDC) is that it enables the investigation of intrinsic characteristics, such as the nonlinearities inherent to this class of materials. Moreover, understanding of the nonlinear response of bulk TMDCs potentially facilitates a multitude of technological applications, as nonlinear optical processes are a fundamental aspect of commercially available frequency mixing and conversion or lasing systems.[33, 34] Harmonic generation offers vast research potential, exemplified by applications such as biosensing [35], microscopy [36] and, more generally, the ability to generate optical pulses varying on attosecond timescales, enabling the study of ultrafast processes with unprecedented temporal and spatial resolution.[37]

Here, we experimentally demonstrate chiral light-matter coupling by merging maximally chiral metasurfaces with excitons in van der Waals (vdW) semiconductors. We develop and experimentally realize a monolithic $WS_2$ metasurface with out-of-plane symmetry breaking, simultaneously addressing two fundamental constraints of cavity physics. First, our approach allows for facile and substantial tuning of the metasurface chiral resonance via the incidence angle, going beyond the established invasive and irreversible cavity tuning via stimuli-responsive materials.[38] Crucially, our tuning mechanism maintains full system functionality (i.e., resonance modulation and maximal chirality) throughout the tuning range, greatly exceeding previous approaches.[39] Second, we experimentally realize a chiral polaritonic system using our vdW metasurface as an open cavity, showing clear handedness-selective anti-crossing behavior and Rabi splitting via self-hybridization of chiral exciton-polaritons. Furthermore, we utilize this system to implement the theoretically predicted but so far experimentally elusive polariton-driven chiral harmonic generation in TMDCs.

# Results

**Maximally Chiral Metasurfaces for Self-Hybridized Chiral Exciton-Polaritons**

Our metasurface design adopts a rod type unit cell geometry with broken out-of-plane symmetry, where each unit cell contains two identical rods lying on different facets (Figs. 1a and 1d). By controlling the opening angle $\alpha$ between the rods and their height difference $\Delta h$, the anti-parallel electric dipole BIC is transformed into a radiative maximally chiral quasi-BIC (qBIC) (see SI Note 1). This approach has previously been shown for chiral metasurfaces at microwave wavelengths using ceramic resonators [40] and close to the red part of the visible spectrum using silicon [41], but has so far not been brought to experiments involving van der Waals materials and wavelengths in the visible range. In our metasurface structure, the resonators are composed solely of bulk $WS_2$ to enable self-hybridized light-matter coupling [42]. In order to realize such structures experimentally, a new fabrication workflow relying on a multi-step top-down nanofabrication process had to be developed, merging both inverse and lift-off techniques. Notably, this method represents a significant improvement over previous approaches because it is not limited to the structuring of sputtered/evaporated materials, thus offering general applicability as evidenced by our work with exfoliated materials, while also achieving superior accuracy (for fabrication details see Methods, SI Note 3).

The qBIC resonance can be spectrally shifted by continuously varying the geometric parameters of the structure. However, since the height of a single flake cannot be continuously tuned, we employ an in-plane scaling factor $S$, which scales the structures only in the x-y-plane. Throughout this work, we utilize left-handed metasurface structures, where the qBIC resonance is uncoupled from right circularly polarized (RCP) light. We observe maximal chirality in experimental transmittance measurements, characterized by negligible cross-polarized signals $T_{RL}$ and $T_{LR}$ and selective mode formation in the co-polarized signal $T_{LL}$ (Fig. 1b). Here, the indices f and i in $T_{fi}$ correspond to the final and initial polarization states, respectively.

The qBIC-driven metasurface concept enables direct control over the radiative quality (Q) factor of the chiral qBIC via introducing the perturbation parameters: opening angle $\alpha$ and height difference $\Delta h$.[41] However, an upper limit on the achievable Q-factor is imposed by material-intrinsic loss channels, as well as other parasitic losses such as fabrication imperfections, surface roughness or edge effects.[43] The perturbation parameters $\alpha$ and $\Delta h$ have a direct impact on the chiral response, which approaches maximal chirality at $\alpha \sim k\Delta h$ (with k being the free-space light wavenumber, SI Note 1).[40] Previous work has shown that the absorbance of the polariton branches in the strong coupling regime is maximized when the intrinsic and radiative damping rates (of the exciton and qBIC, respectively) in the system are matched, a phenomenon known as polaritonic critical coupling.[42] This optimal absorbance into the polaritonic branches decreases for qBIC linewidths that either are broader or narrower than the exciton linewidth.

Even though maximally chiral resonances with higher Q-factors can be achieved in our system through simultaneous decrease in the opening angle $\alpha$ and height difference $\Delta h$, this would move the metasurface from the polaritonic critical coupling regime and make the resonances more prone to collapse due to higher intrinsic material losses in the upper polariton band (UPB). Consequently, we deliberately selected broader resonances for the experiment (opening angle $\alpha = 12°$ and $\Delta h = 70$ nm, Fig. 1b), as they allowed to resolve the UPB with negligible impact on coupling strength or Rabi splitting, while also maximizing light absorption into the polaritonic branches.

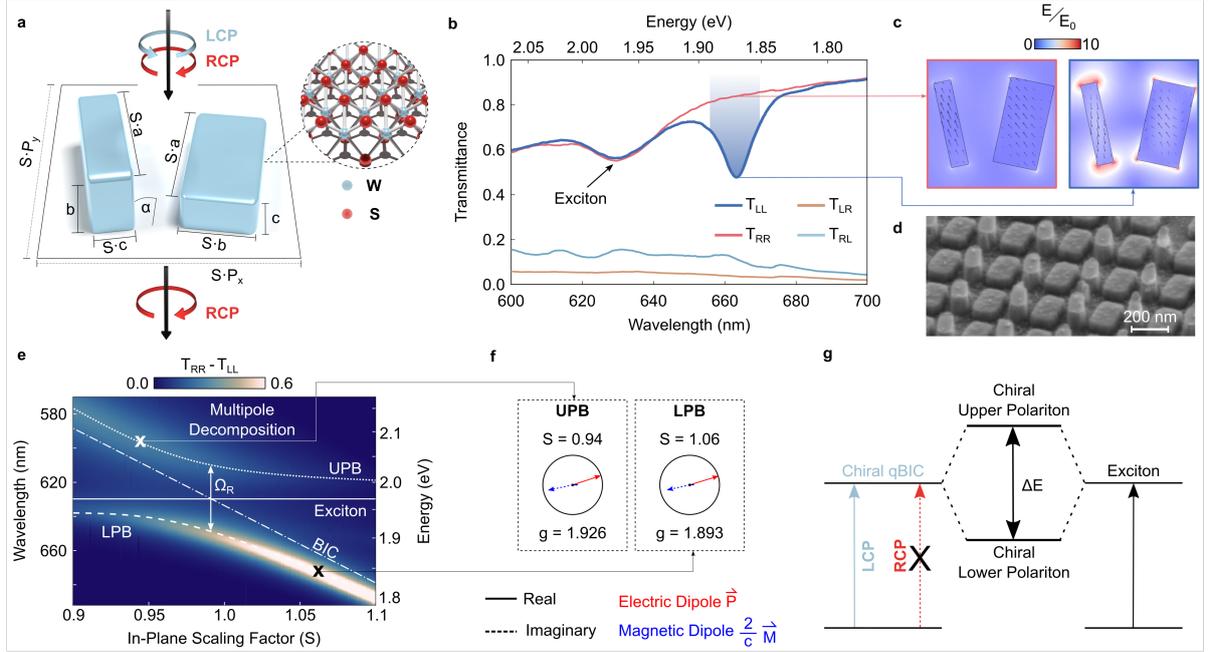

**Figure 1: Maximally chiral WS$_2$ metasurfaces for strong coupling**. **a** Out-of-plane symmetry breaking enables chiral qBIC resonances in three-dimensional monolithic WS$_2$ metasurfaces. The unit cell design has a periodicity of $P_x = P_y = 375$ nm, with rod side lengths of a = 225 nm, b = 115 nm, and c = 45 nm, respectively. The height difference is $\Delta h$ = b - c = 70 nm. Each rod is rotated by an angle of $\alpha = 12°$. **b** Experimental co-/cross-polarized chiral transmittance spectra verify maximal chirality, showing the selective emergence of a qBIC resonance in the left co-polarized signal. **c** Simulated electric near-field enhancement (z = 80 nm, i.e. average height value of the rods) at the qBIC resonance for a left-handed structure under LCP/RCP illumination. **d** SEM image of the fabricated chiral metasurface on a SiO$_2$ substrate. **e** Simulated transmittance difference $\Delta T=T_{RR}-T_{LL}$ of the metasurface for different in-plane scaling factors S shows the emergence of lower (LPB) and upper (UPB) polaritonic branches with anticrossing behavior. **f** Multipole decomposition analysis of the eigenstates of two different scaling factors, one from each of the respective polariton branches. The dissymmetry factor $g$ (see Eq. (1)) reaches values close to the theoretical upper limit $g = 2$ for both cases. **g** Simplified energy level diagram of self-hybridization of excitons and qBIC into chiral polaritons.

The simulated electric near field at resonance exhibits no enhancement for RCP excitation, whereas a tenfold enhancement is observed for left circularly polarized (LCP) excitation, verifying the formation of a maximally chiral mode (Fig. 1c). By varying the geometry of the metasurface through the in-plane scaling factor S, the qBIC resonance wavelength can be shifted across the spectral range of the room-temperature exciton in $WS_2$ at 629 nm (1.971 eV). The resulting transmittance difference $\Delta T = T_{RR} - T_{LL}$ exhibits a characteristic anticrossing pattern around the exciton position (Fig. 1e).

To assess the chirality of the resulting exciton-polaritons, we numerically perform multipole decomposition of the hybrid eigenstates. Specifically, we decompose the coupling coefficients between the displacement current associated with the polaritonic eigenstate and free-space plane waves for metasurfaces with a given geometric scaling. The computed dipole moments of these eigenstates (Fig. 1f), closely satisfy the condition for maximal chirality (SI Note 2). To quantify the chiral nature of our polaritons and draw an analogy with molecular enantiomers we evaluate the dissymmetry factor $g$, defined as the ratio of rotational strength $R$ to dipole strength $D$ [44]:

$$g = \frac{4R}{D} = \frac{4\,\text{Im}(2\mathbf{M} \cdot c\mathbf{P})}{|2\mathbf{M}|^2 + |c\mathbf{P}|^2} \quad (1)$$

where $\mathbf{P}$ and $\mathbf{M}$ denote the electric and magnetic dipole moments, respectively. Since we consider polaritons confined within the bulk meta-atoms, whose sizes are much larger than those of typical molecular enantiomers, we cannot neglect the contribution of the electric quadrupole, which is of the same order in the multipole expansion as the magnetic dipole. This results in an extra doubling of $\mathbf{M}$ in (Eq. 1) (See SI Note 2 for more details). The numerically calculated values of $g$ for two representative scaling factors on both polariton branches approach the theoretical upper limit of $g = 2$, confirming that the resulting polaritonic eigenstates very closely approach the condition of maximum optical chirality. The energy level diagram for this process in a left-handed metasurface is shown in Fig. 1g (for full derivation see SI Note 2). This formation of self-hybridized chiral exciton-polaritons highlights our metasurface as a promising platform for chiral polaritonics.

**Chiral Strong Coupling in k-Space**

Although the concept of qBICs in metasurfaces composed of nanorods with differing heights and antiparallel dipole configurations has been previously demonstrated[40, 41], the k-space characteristics of such systems have remained unexplored. Here, we analyze the far-field polarization of the eigenstates supported by the metasurface structure depicted in Fig. 1 (with in-plane scaling factor $S = 1$ and, for more clarity, we neglect refractive index frequency dispersion and set $n = 4.5$.) As shown in Fig. 2a, the mode preserves its circular polarization throughout the entire k-space region under consideration, demonstrating significantly improved robustness against variations in the angle of incidence compared to previously reported systems.[39]

Concurrently, the resonance wavelength of the chiral qBIC exhibits a saddle-shaped dispersion as a function of in-plane wavevector $\mathbf{k}$, redshifting with increasing $|k_x|$ (Fig. 2b). This behavior enables continuous spectral tuning of the resonance position. Notably, the Q-factor increases with $|k_x|$ but remains within experimentally feasible values, not exceeding 120 (Fig. 2c). Therefore, the angle of incidence no longer acts as a constraint on a system, but rather as a post-fabrication degree of freedom for sub-nanometer precise resonance wavelength control and tuning. By varying the angle of incidence, the resonance wavelength of the chiral qBIC can be precisely tuned over a spectral range exceeding 50 nm. This allows for spectral shift across with the excitonic resonance of WS$_2$ at 629 nm (Fig. 2b), thus enabling to probe the chiral strong coupling using a single metasurface.

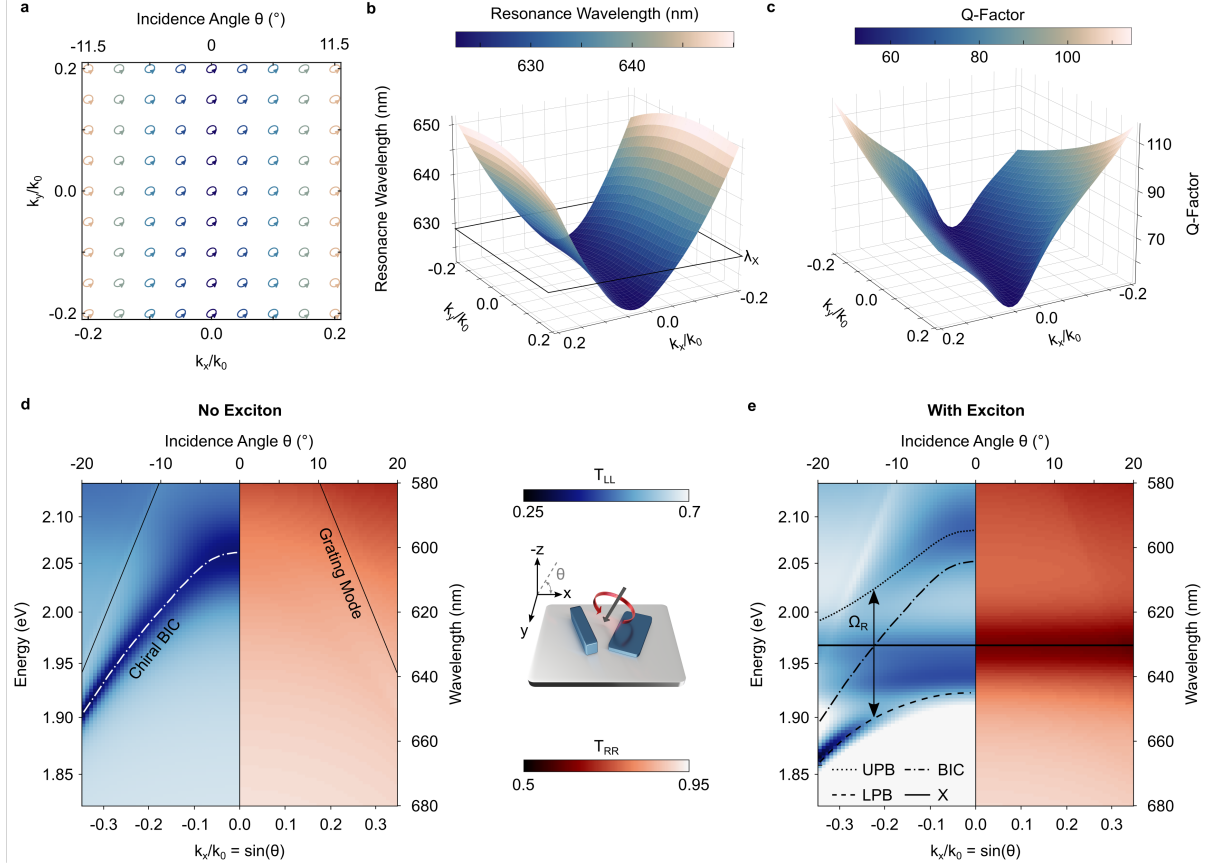

**Figure 2: Angular dispersion analysis of chiral strong coupling**. **a** The simulated far-field polarization of the left-handed metasurface confirms that the chiral qBIC's maximum chirality remains unaffected by oblique incidence. **b** Simulated dispersion for the left-handed chiral qBIC resonance in k-space shows a saddle-shaped resonance wavelength. **c** Q-factor tuning behavior for varying incidence angle $\theta$. The general saddle shape for Q-factor shows an additional narrowing of the line width when $k_x$ and $k_y$ have opposite signs. **d** Simulated co-polarized LCP (left) and RCP (right) transmittance spectra of the left-handed WS$_2$ metasurface for different incidence angles $\theta$ in the yz-plane and without the exciton. The chiral qBIC sustains its polarization when moving to oblique incidence, enabling post-fabrication tuning of the resonance wavelength via the incidence angle. **e** Simulated co-polarized transmission of LCP (left) and RCP (right) light for different incidence angles $\theta$ of the metasurface with the exciton. LCP spectra show anticrossing formation while only the constant excitonic fingerprint is visible for RCP excitation.

To illustrate this effect, we simulate the co-polarized transmission spectra $T_{RR}$ and $T_{LL}$ as a function of $k_x$ (with $k_y = 0$), using a metasurface geometry with $S = 0.95$ to shift the normal incidence qBIC resonance wavelength toward 600 nm. In the absence of the excitonic response (material model defined in SI Note 1), the chiral qBIC mode appears exclusively in $T_{LL}$ and shifts to longer wavelengths with increasing $k_x$ (Fig. 2d). Even at large in-plane momentum values ($k_x = 0.34$, corresponding to an incidence angle $\theta = 20°$), the mode maintains its chiral character. Incorporating the excitonic contribution into the permittivity model leads to the formation of two polaritonic branches, clearly resolved in the $T_{LL}$ spectra (Fig. 2e), using only a single metasurface. In contrast, $T_{RR}$ reveals no hybridized resonances and only a weak signature of the exciton. This behavior highlights the metasurface as a continuously tunable photonic platform, enabling non-invasive and reversible resonance control without the need for externally responsive materials, in contrast to conventional cavity systems.[38, 45]

**Experimental Chiral Strong Coupling**

To demonstrate the formation of chiral self-hybridized exciton polaritons in our experiments, we fabricate a series of chiral $WS_2$ metasurfaces with a range of in-plane scaling factors from $S = 0.97$ to $S = 1.09$. Analysis of SEM images gave the following dimensions for the structure (for $S = 1$): widths of 130 nm and 63 nm, heights of 80 nm and 30 nm, lengths of 210 nm and a periodicity of 370 nm, which also were confirmed by post-fabricational simulations (SI Note 4). The corresponding circular dichroism CD = $(T_{RR}-T_{LL})/(T_{RR}+T_{LL})$ spectra show a clear anticrossing behavior - a preliminary indicator of strong coupling (Fig. 3a).

We employ a temporal coupled mode theory (TCMT) model to fit the transmittance spectra (see SI Note 5), allowing us to rigorously assess the coupling strength. Unlike conventional polariton dispersion analysis, this method provides direct access to both the dispersion and linewidth of the underlying qBIC, offering a more complete and quantitative characterization of the coupled system. The linewidth of the $WS_2$ exciton is $\hbar\gamma_X = 36$ meV. This value, as well as the excitonic energy at $\hbar\omega_X = 1.971$ eV, are kept fixed for the TCMT fits used to extract the coupling strength. We further assume that the scaling factor and the qBIC dispersion follow a linear relationship (Fig. 3b), a common approximation in literature [46]. The Rabi energy is defined as

$$\hbar\Omega_R = 2\hbar\sqrt{\kappa^2 - \left(\frac{\gamma_{\text{BIC}} - \gamma_{\text{X}}}{4}\right)^2}, \quad (2)$$

where $\Omega_R$ denotes the Rabi frequency and $\kappa$ is the coupling strength. Based on our TCMT fitting, we obtain a Rabi splitting of $\hbar\Omega_R = 108$ meV.

To verify that the system operates in the strong coupling regime, it is essential to assess whether the rate of coherent energy exchange between the chiral qBIC and the exciton mode exceeds the rate at which each resonance dissipates energy into its respective loss channels. This condition ensures that the light-matter interaction gives rise to hybridized polariton modes with well-resolved, spectrally separable features. The first commonly used criterion $c_1$ compares the Rabi frequency $\Omega_R$ with the sum of the linewidths of the exciton and qBIC modes [47]:

$$c_1 = \frac{\Omega_R}{\gamma_{\text{BIC}} + \gamma_{\text{X}}} > 1, \tag{3}$$

indicating that the polariton mode splitting should exceed the total dissipation rate of the uncoupled modes. For our system, the calculated $c_1 = 1.5$ clearly satisfies this criterion. A more stringent condition is given by the second criterion $c_2$, which accounts for unequal linewidths and compares the coupling strength $\kappa$ with the root-mean-square average of the losses [47]:

$$c_2 = \frac{\kappa}{\sqrt{\frac{\gamma_{\text{X}}^2 + \gamma_{\text{BIC}}^2}{2}}} > 1. \tag{4}$$

Our system also meets this stricter condition, with a calculated $c_2 = 1.5$. Combining this with the observation of distinct anticrossing behavior, our metasurface is clearly within the chiral strong coupling regime.

The TCMT analysis further reveals that the system can be tuned towards the polaritonic critical coupling regime via straightforward geometrical variations. This regime is characterized by balanced linewidths of both exciton and chiral qBIC, leading to optimal energy transfer between the excitonic and photonic components of the hybrid mode (Fig. 3b). At normal incidence ($k_x = 0$), different scaling factors exhibit the expected chiral response originating from the hybridization of maximally chiral qBIC and exciton: the left-handed signal $T_{\text{LL}}$ shows pronounced UPB and LPB dips, whereas the right-handed trace $T_{\text{RR}}$ shows only the bare excitonic feature (Fig. 3c). This further confirms that the chiral qBIC continues to couple selectively to LCP even as the exciton–photon detuning is varied via geometric scaling.

To probe the chiral character of the exciton-polariton modes within a particular metasurface, we leverage the previously established circular polarization stability in k-space of our design to perform angle-resolved reflectance measurements under circularly polarized illumination (see Methods), scanning the in-plane momentum $k_x = \sin\theta$, where $\theta$ is the angle of reflectance. Spectra were recorded separately for LCP and RCP light ($R_{\text{LL}}$ and $R_{\text{RR}}$) using back focal plane spectroscopy. The resulting dispersion in $R_{\text{LL}}$ reveals clear signatures of strong coupling between the exciton and the chiral qBIC photonic mode, giving rise to UPB and LPB with a Rabi splitting that emerges at finite in-plane momentum (see SI Note 6). In contrast, apart from a faint background that mirrors the bare excitonic absorption, no resonant chiral features are observed in $R_{\text{RR}}$ (Fig. 3d).

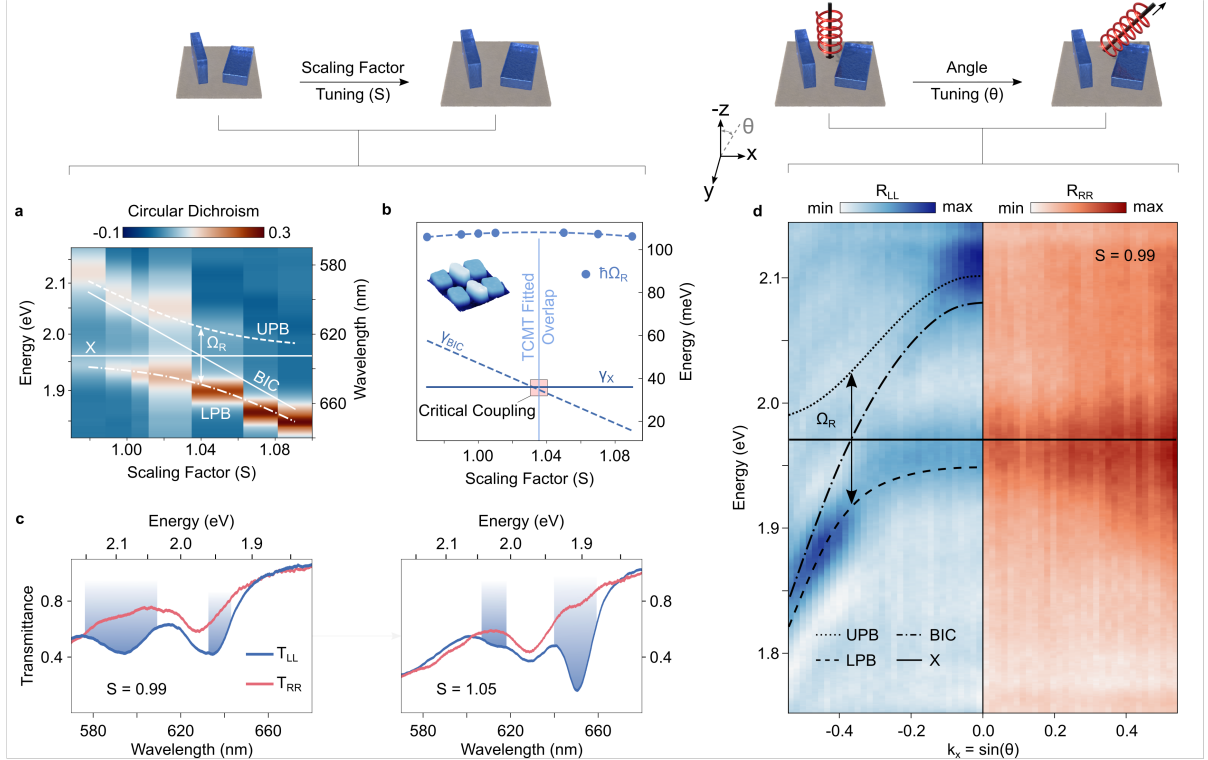

**Figure 3: Experimental chiral strong coupling in real and k-space. a** Tuning of the chiral qBIC resonance via scaling factor S results in polaritonic splitting into upper and lower branches. The dashed lines show the polariton dispersion extracted via TCMT fitting, yielding a Rabi splitting of 108.06 meV. **b** Extracted linewidths for chiral qBIC and exciton from TCMT fits, along with the fitted Rabi energy. At the overlap position, the fitted linewidths of exciton and chiral qBIC are close to becoming equal, indicating that the system approaches the regime of polaritonic critical coupling, the point of maximal energy exchange between light and matter. The inset shows an AFM image of the metasurface structure. **c** Transmittance spectra under LCP and RCP excitation for different scaling factor geometries at normal incidence. **d** Experimental back focal plane imaging of the co-polarized reflectance spectra $R_{LL}$ and $R_{RR}$. Dashed lines highlight the polaritonic dispersion in left-handed signal ($S = 0.99$).

Side-by-side comparison of the two dispersions reveals the momentum-resolved "fingerprint" of a metasurface whose optical response depends on the handedness of the incident light. The engineered chiral qBIC serves as a selective interface between photons and excitons, enabling efficient self-hybridization of chiral exciton-polaritons for LCP while remaining passive for RCP.

## Chiral Third Harmonic Generation in the Strong Coupling Regime

In centrosymmetric bulk $WS_2$, third-harmonic generation (THG) provides a nonlinear optical window into electronic and excitonic properties. When strong light-matter coupling is realized, the formation of exciton–polaritons modifies the optical response, including in the nonlinear regime. By resolving THG under circularly and linearly polarized excitation (Fig. 4a and c), the polarization-dependent signatures of these hybrid states can be accessed. Our nonlinear measurements on unpatterned WS$_2$ flakes reveal a pronounced intrinsic enhancement of nonlinearities near the exciton resonance (Fig. 4b, see SI Note 7 and 8). Outside this excitonic region, the THG intensity decreases by more than an order of magnitude, corresponding to a reduction of the nonlinear susceptibility by over a factor of three.

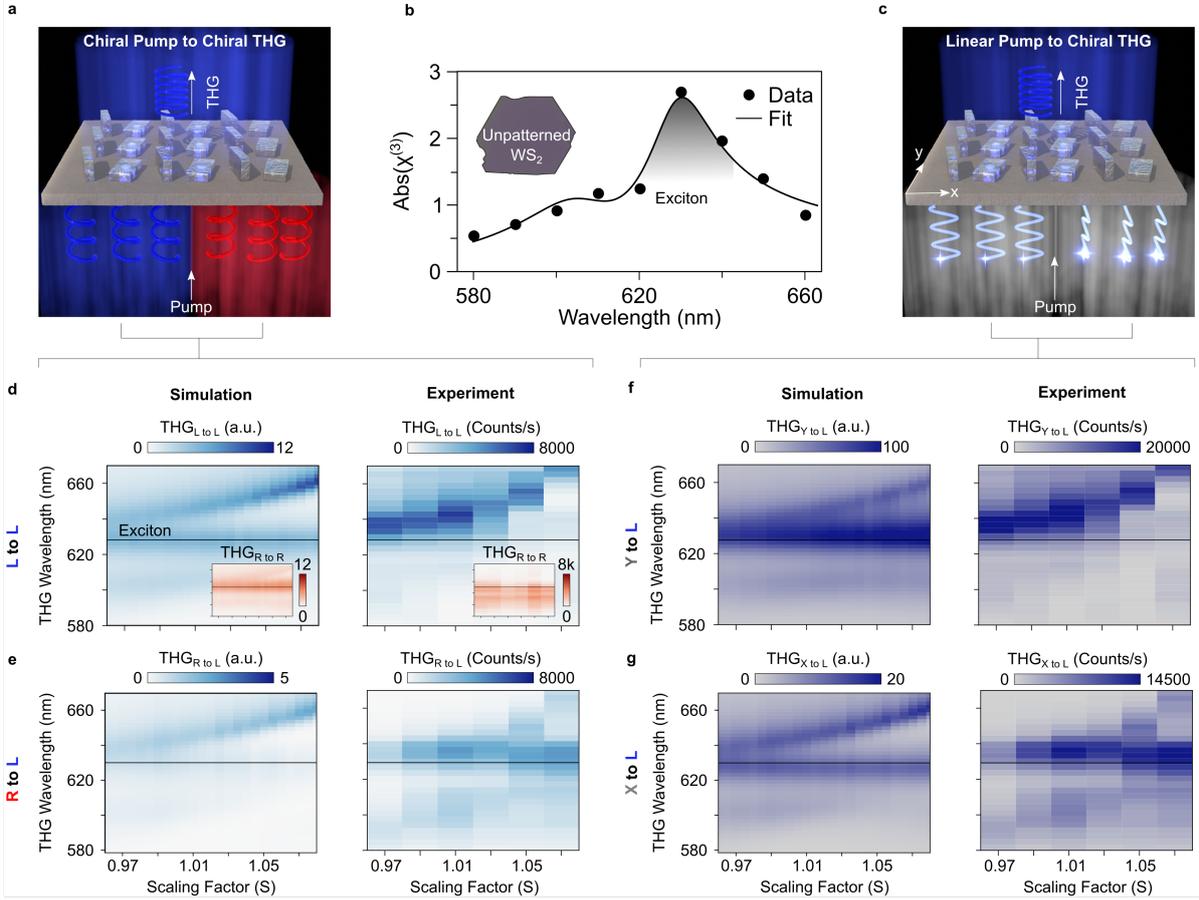

Figure 4: **Chiral exciton-polariton nature of third harmonic generation.** **a** Conceptual visualization of the chiral pumping harmonic generation experiment, where the emission is resonant with the chiral qBIC. **b** Nonlinear susceptibility $\chi^{(3)}$ of unpatterned bulk WS$_2$ extracted from experimental data, showing a native excitonic nonlinear enhancement. **c** Conceptual visualization of the linear pumping driven chiral harmonic generation experiment, again with emission resonant with the chiral qBIC. **d** Simulated and experimental left co-polarized THG, exhibiting hybridization of the exciton and the THG. The insets show right co-polarized THG only displaying native excitonic signal. **e** Right to left cross-polarized THG again displays hybridization of the exciton and the THG. The simulated and experimental left handed THG from linear pump, also exhibits hybridization of the exciton and the THG despite the linear pump polarization (along **f** $Y$ or **g** $X$).

In most qBIC-based nonlinear studies, the pump is tuned to the qBIC resonance so that the local field enhancement amplifies the material's nonlinear response. Here, however, the pump is is tuned between 1740 nm and 2000 nm in steps of 20 nm making the third-harmonic emission itself occur in the resonant wavelength range. This choice is dictated by the fundamental optical and electronic properties of the material: pumping at the exciton wavelength would shift the third harmonic into the ultraviolet, where WS$_2$ is strongly absorbing. Comparing chiral THG intensities from metasurfaces with different scaling factors reveals that the handedness-dependent hybridization between the chiral qBIC and the exciton also governs the nonlinear response (Fig. 4d). Here, the strong third-order nonlinearity of the WS$_2$ exciton couples to the chiral qBIC, giving rise to hybrid chiral polaritons that act as nonlinear emitters. This hybridization inherits the chiral character of the photonic mode and the nonlinear enhancement of the exciton, effectively merging their respective advantages into a single polaritonic response.

To further explore the origin of this nonlinear chirality, we must consider the field distribution within the metasurface, which results from the hybridization of the chiral qBIC mode with the exciton. The resulting chiral polaritons act as nanoscale chiral point sources. When the left-handed metasurface is pumped by RCP light the generated nonlinear emission is governed not by the polarization state of the incident light, but by the intrinsic handedness of the chiral polariton. Consequently, even RCP excitation yields a LCP third-harmonic signal (Fig. 4e). Only the excitonic response can be observed in the corresponding RCP third-harmonic signal (Fig. 4d).

To show that the effect is not limited to circularly polarized pumping, we measured THG under linearly $X$- and $Y$-polarized excitation (Fig. 4f,g). Simulations and experiments alike reveal a pronounced polariton-induced splitting in the LCP third-harmonic channel, while the RCP signal remains weak (see SI Note 9). This again indicates that the observed nonlinear chirality is imposed by the polariton, not by the pump polarization or any intrinsic material chirality. Transforming a purely linear pump into helicity-controlled harmonic emission within a passive, CMOS-compatible dielectric metasurface opens several avenues. This control arises not from the light source or material chirality, but from chiral polaritons—hybrid light-matter states with geometry-defined handedness. The metasurface design, parameterized by an opening angle $\alpha$, enables compact, ultrafast sources of circularly polarized light for chiral spectroscopy and valley-selective 2D material control. Left- and right-handed elements can be integrated for helicity-multiplexed channels without active optics. Additionally, polariton-induced spectral splitting offers tunable frequency control for nonlinear signal processing. This approach connects chiral photonics with polariton physics for on-chip, helicity-engineered light sources.

## Discussion

We have developed an out-of-plane symmetry-broken metasurface fabricated from bulk TMDC to demonstrate and experimentally verify, for the first time to our knowledge, chiral self-hybridized exciton-polaritons, pushing beyond two major bottlenecks of cavity-polariton physics. Firstly, the generation of chiral fields is unattainable in conventional closed Fabry-Pérot cavity systems because of phase-flipping at mirror interfaces.[48] The need for a handedness-preserving mirror has already been understood and experimental proof-of-concept metamaterial-designs were proposed in the GHz range [49] as well as theoretically investigated in the visible range [24]. For the latter theoretical study, handedness-preservation was achieved by hybridization of different parity eigenstates. However, experimental realizations of more general chiral polaritonic cavity approaches have remained elusive.

By introducing the physics of metasurfaces with maximum chiral qBICs, we elevate the concept of chiral cavities—realizing a platform that preserves the handedness of a localized high-Q optical mode in the visible range, while offering exceptional design flexibility and dynamic tunability. As a result, we observe the formation of chiral self-hybridized exciton-polaritons. Secondly, the metasurface resonance is tunable over a substantial range by using the angle of incidence as noninvasive post-fabrication tuning mechanism without adversely impacting the chiral performance. This presents a step forward for polaritonic physics, as it provides a clear pathway towards not only chiral cavities but also greatly simplified detuning without the need for highly specialized stimuli-responsive materials [50] or complex electrically driven designs [45].

By leveraging the out-of-plane symmetry breaking of our chiral qBIC metasurface, we reveal the previously unobserved polariton-driven chararcteristics of THG in a bulk TMDC. The nonlinear process is mediated by the chiral qBIC and hybridizes with the A-exciton, yielding a spectrally split, circularly polarized THG output, even under a linearly polarized pump. This control over bulk nonlinearities via exciton-polariton opens avenues for chip-scale quantum light emitters [51], ultrafast spin-selective light sources [52], and reconfigurable nonlinear mirrors [53]. Because the platform uses bulk vdW materials and high-quality centimeter-scale TMDC films can now be routinely obtained by vapor deposition [54], the architecture is highly scalable. In particular, its ability to translate widely available linearly polarized laser inputs into helicity-programmed, frequency-converted outputs makes it an attractive candidate for integrated optical parametric generation and amplification in advanced photonic circuits.[55] Through our generally applicable chiral metasurface design, the same strategy can be extended to material systems that exhibit strong intrinsic polaritonic or nonlinear responses, including halide perovskites, magnetic van der Waals crystals such as CrSBr, and Moiré-twisted heterostructures. Implementing chirality-driven symmetry breaking in these emerging quantum materials could enable tailored polariton landscapes, valley-specific nonlinearities and other symmetry-protected phenomena, paving the way toward a new class of quantum-engineered photonic devices.

# References

1. Zhang, C. *et al.* Quantum plasmonics pushes chiral sensing limit to single molecules: a paradigm for chiral biodetections. *Nature Communications* **15,** 2 (2024).

2. Garcia-Guirado, J., Svedendahl, M., Puigdollers, J. & Quidant, R. Enhanced chiral sensing with dielectric nanoresonators. *Nano Letters* **20,** 585–591 (2019).

3. Li, X. *et al.* Proximity-induced chiral quantum light generation in strain-engineered WSe2/NiPS3 heterostructures. *Nature Materials* **22,** 1311–1316 (2023).

4. Shreiner, R., Hao, K., Butcher, A. & High, A. A. Electrically controllable chirality in a nanophotonic interface with a two-dimensional semiconductor. *Nature Photonics* **16,** 330–336 (2022).

5. Claassen, M., Jia, C., Moritz, B. & Devereaux, T. P. All-optical materials design of chiral edge modes in transition-metal dichalcogenides. *Nature Communications* **7,** 13074 (2016).

6. Wang, G. *et al.* Colloquium: Excitons in atomically thin transition metal dichalcogenides. *Reviews of Modern Physics* **90,** 021001 (2018).

7. Chen, P. *et al.* Chiral coupling of valley excitons and light through photonic spin–orbit interactions. *Advanced Optical Materials* **8,** 1901233 (2020).

8. Zibouche, N., Philipsen, P., Kuc, A. & Heine, T. Transition-metal dichalcogenide bilayers: Switching materials for spintronic and valleytronic applications. *Physical Review B* **90,** 125440 (2014).

9. Lee, J. *et al.* Hydrogen evolution reaction at anion vacancy of two-dimensional transition-metal dichalcogenides: ab initio computational screening. *The Journal of Physical Chemistry Letters* **9,** 2049–2055 (2018).

10. Varsano, D., Palummo, M., Molinari, E. & Rontani, M. A monolayer transition-metal dichalcogenide as a topological excitonic insulator. *Nature Nanotechnology* **15,** 367–372 (2020).

11. Shi, W. *et al.* Superconductivity series in transition metal dichalcogenides by ionic gating. *Scientific Reports* **5,** 12534 (2015).

12. Schneider, C., Glazov, M. M., Korn, T., Höfling, S. & Urbaszek, B. Two-dimensional semiconductors in the regime of strong light-matter coupling. *Nature Communications* **9,** 2695 (2018).

13. Sanvitto, D. & Kéna-Cohen, S. The road towards polaritonic devices. *Nature Materials* **15,** 1061–1073 (2016).

14. Ghosh, S. & Liew, T. C. Quantum computing with exciton-polariton condensates. *npj Quantum Information* **6,** 16 (2020).

15. Kang, J.-W. *et al.* Room temperature polariton lasing in quantum heterostructure nanocavities. *Science Advances* **5,** eaau9338 (2019).

16. Dufferwiel, S. *et al.* Exciton–polaritons in van der Waals heterostructures embedded in tunable microcavities. *Nature Communications* **6,** 8579 (2015).

17. Baranov, D. G., Schäfer, C. & Gorkunov, M. V. Toward molecular chiral polaritons. *ACS Photonics* **10,** 2440–2455 (2023).

18. Cheng, Q. *et al.* Tuning the plexcitonic optical chirality using discrete structurally chiral plasmonic nanoparticles. *Nano Letters* **23,** 11376–11384 (2023).

19. Zhu, J. *et al.* Strong light–matter interactions in chiral plasmonic–excitonic systems assembled on DNA origami. *Nano Letters* **21,** 3573–3580 (2021).

20. Stamatopoulou, P. E., Droulias, S., Acuna, G. P., Mortensen, N. A. & Tserkezis, C. Reconfigurable chirality with achiral excitonic materials in the strong-coupling regime. *Nanoscale* **14,** 17581–17588 (2022).

21. Wang, X., Ronca, E. & Sentef, M. A. Cavity quantum electrodynamical Chern insulator: Towards light-induced quantized anomalous Hall effect in graphene. *Physical Review B* **99,** 235156 (2019).

22. Frisk Kockum, A., Miranowicz, A., De Liberato, S., Savasta, S. & Nori, F. Ultrastrong coupling between light and matter. *Nature Reviews Physics* **1,** 19–40 (2019).

23. Hübener, H. *et al.* Engineering quantum materials with chiral optical cavities. *Nature Materials* **20,** 438–442 (2021).

24. Voronin, K., Taradin, A. S., Gorkunov, M. V. & Baranov, D. G. Single-handedness chiral optical cavities. *ACS Photonics* **9,** 2652–2659 (2022).

25. Yu, N. & Capasso, F. Flat optics with designer metasurfaces. *Nature Materials* **13,** 139–150 (2014).

26. Valentine, J. *et al.* Three-dimensional optical metamaterial with a negative refractive index. *Nature* **455,** 376–379 (2008).

27. Yin, X., Ye, Z., Rho, J., Wang, Y. & Zhang, X. Photonic spin Hall effect at metasurfaces. *Science* **339,** 1405–1407 (2013).

28. Koshelev, K., Bogdanov, A. & Kivshar, Y. Meta-optics and bound states in the continuum. *Science Bulletin* **64,** 836–842 (2019).

29. Gorkunov, M. V., Antonov, A. A. & Kivshar, Y. S. Metasurfaces with maximum chirality empowered by bound states in the continuum. *Physical Review Letters* **125,** 093903 (2020).

30. Zhou, H. *et al.* Photonic spin-controlled self-hybridized exciton-polaritons in WS 2 metasurfaces driven by chiral quasibound states in the continuum. *Physical Review B* **109,** 125201 (2024).

31. Löchner, F. J. *et al.* Hybrid dielectric metasurfaces for enhancing second-harmonic generation in chemical vapor deposition grown MoS2 monolayers. *ACS Photonics* **8,** 218–227 (2020).

32. Sortino, L. *et al.* Dielectric nanoantennas for strain engineering in atomically thin two-dimensional semiconductors. *ACS Photonics* **7,** 2413–2422 (2020).

33. Pu, Y., Grange, R., Hsieh, C.-L. & Psaltis, D. Nonlinear Optical Properties of Core-Shell Nanocavities for Enhanced Second-Harmonic Generation. *Physical Review Letters* **104,** 207402 (2010).

34. Brabec, T. & Krausz, F. Intense few-cycle laser fields: Frontiers of nonlinear optics. *Reviews of Modern Physics* **72,** 545 (2000).

35. Tran, R. J., Sly, K. L. & Conboy, J. C. Applications of surface second harmonic generation in biological sensing. *Annual Review of Analytical Chemistry* **10,** 387–414 (2017).

36. James, D. S. & Campagnola, P. J. Recent advancements in optical harmonic generation microscopy: Applications and perspectives. *BME Frontiers* (2021).

37. Li, J. *et al.* Attosecond science based on high harmonic generation from gases and solids. *Nature Communications* **11,** 2748 (2020).

38. Hennessy, K., Högerle, C., Hu, E., Badolato, A. & Imamoğlu, A. Tuning photonic nanocavities by atomic force microscope nano-oxidation. *Applied Physics Letters* **89** (2006).

39. Choi, M., Alù, A. & Majumdar, A. Observation of photonic chiral flatbands. *Physical Review Letters* **134,** 103801 (2025).

40. Gorkunov, M. V., Antonov, A. A., Tuz, V. R., Kupriianov, A. S. & Kivshar, Y. S. Bound states in the continuum underpin near-lossless maximum chirality in dielectric metasurfaces. *Advanced Optical Materials* **9,** 2100797 (2021).

41. Kühner, L. *et al.* Unlocking the out-of-plane dimension for photonic bound states in the continuum to achieve maximum optical chirality. *Light: Science & Applications* **12,** 250 (2023).

42. Weber, T. *et al.* Intrinsic strong light-matter coupling with self-hybridized bound states in the continuum in van der Waals metasurfaces. *Nature Materials* **22,** 970–976 (2023).

43. Bernhardt, N. *et al.* Quasi-BIC resonant enhancement of second-harmonic generation in WS2 monolayers. *Nano Letters* **20,** 5309–5314 (2020).

44. Condon, E. U. Theories of optical rotatory power. *Reviews of Modern Physics* **9,** 432 (1937).

45. He, X. *et al.* Electrically driven highly tunable cavity plasmons. *ACS Photonics* **6,** 823–829 (2019).

46. Nan, L. *et al.* Angular dispersion suppression in deeply subwavelength phonon polariton bound states in the continuum metasurfaces. *Nature Photonics,* 1–9 (2025).

47. Cao, S. *et al.* Normal-incidence-excited strong coupling between excitons and symmetry-protected quasi-bound states in the continuum in silicon nitride–WS2 heterostructures at room temperature. *The Journal of Physical Chemistry Letters* **11,** 4631–4638 (2020).

48. Plum, E. & Zheludev, N. I. Chiral mirrors. *Applied Physics Letters* **106** (2015).

49. Fedotov, V., Rogacheva, A., Zheludev, N., Mladyonov, P. & Prosvirnin, S. Mirror that does not change the phase of reflected waves. *Applied Physics Letters* **88** (2006).

50. Kim, J. *et al.* Dynamic control of nanocavities with tunable metal oxides. *Nano Letters* **18,** 740–746 (2018).

51. Chakraborty, C., Vamivakas, N. & Englund, D. Advances in quantum light emission from 2D materials. *Nanophotonics* **8,** 2017–2032 (2019).

52. Garmire, E. Nonlinear optics in daily life. *Optics Express* **21,** 30532–30544 (2013).

53. Stankov, K. & Jethwa, J. A new mode-locking technique using a nonlinear mirror. *Optics Communications* **66,** 41–46 (1988).

54. Shen, F. *et al.* Transition metal dichalcogenide metaphotonic and self-coupled polaritonic platform grown by chemical vapor deposition. *Nature Communications* **13,** 5597 (2022).

55. Keller, U. Recent developments in compact ultrafast lasers. *Nature* **424,** 831–838 (2003).

56. Munkhbat, B., Wróbel, P., Antosiewicz, T. J. & Shegai, T. O. Optical constants of several multilayer transition metal dichalcogenides measured by spectroscopic ellipsometry in the 300–1700 nm range: high index, anisotropy, and hyperbolicity. *ACS Photonics* **9,** 2398–2407 (2022).

57. Kim, S. *et al.* Chiral electroluminescence from thin-film perovskite metacavities. *Science Advances* **9,** eadh0414 (2023).

## Methods

**Numerical simulations**

The refractive index of the $SiO_2$ substrate was set as 1.45, while that of the $WS_2$ rods was taken from literature.[56] Simulations of transmittance spectra for the 3D-chiral $WS_2$ metasurfaces were conducted using CST Studio Suite 2021 with periodic Floquet boundary conditions. Farfield polarization and transmittance behavior under oblique incidence were numerically investigated using the Electromagnetic Waves Frequency Domain module of COMSOL Multiphysics in 3D mode using a previously developed approach.[57] The tetrahedral spatial mesh for FEM was automatically generated by COMSOL's physics-controlled preset. Simulations were performed within a rectangular spatial domain containing a single metasurface unit cell with periodic boundary conditions applied to its sides. Circularly polarized ports were set at the top and bottom to simulate excitation and registration. To simulate THG two problems at fundamental and third harmonic wavelengths were solved simultaneously using COMSOL Multiphysics. To connect two studies a polarization node was set up with the fields of the linear problem in the volume of the rods.

**Sample fabrication**

Fused silica substrates were initially cleaned by sonication in Acetone at 55°C, followed by Isopropanol to remove any residual Acetone. Subsequently, the substrates were treated with $O_2$ plasma to eliminate organic residue and enhance flake adhesion. To facilitate precise global alignment of the flake position on the substrate during subsequent processing, a marker system was created on the substrates using optical lithography (SÜSS Maskaligner MA6). $WS_2$ flakes were mechanically exfoliated from bulk crystals (HQ Graphene) onto the cleaned silica marker substrates. The deposition process was conducted at a temperature of 105°C to evaporate moisture and stretch the exfoliation tape, ensuring flattened transferred flakes. The height of the flakes was measured using a profilometer (Bruker Dektak XT) with a stylus having a radius of 2 µm. The three-dimensional $WS_2$ metasurfaces were fabricated using a multi-step E-beam lithography (EBL) process, followed by lift-off and reactive-ion etching (RIE). All EBL steps were carried out using an eLINE Plus (Raith Nanofabrication). For the first EBL step, the films were spin-coated with a positive electron beam resist, CSAR 62 (Allresist) and Espacer 300Z. The right half of the unit cell was fully exposed via electron beam lithography using an eLINE Plus system (Raith) at 30 kV with a 15 $\mu$m aperture. The patterns were developed in an amyl acetate bath, followed by a MIBK:IPA (1:9 ratio) bath. The intended height difference was etched into the flakes via RIE (Oxford PlasmaPro 100) using $SF_6$-based chemistry at a pressure of 20 mTorr and an RF power of 50 W using the unexposed resist as a hardmask. Subsequently, a single layer of positive-tone polymethylmethacrylate (PMMA, Kayaku Advanced Materials) was used as the EBL resist, spincoated at 3000 RPM and baked at 180°C for three minutes. Espacer 300Z was then spin-coated onto the sample. The full unit cell design was patterned with the same EBL system at 20 kV with a 15 $\mu$m aperture. Development took place in a solution of 40 ml Ethanol mixed with 7.5 ml of DI Water for 20 seconds. Prior to the first EBL step, a local marker system (50 nm thick Au) was installed in direct proximity to the flake for precise realignment during subsequent steps. After the second patterning run, a hardmask consisting of 2.5 nm titanium, followed by 40 nm of gold was evaporated onto the sample using electron beam evaporation and subsequently lifted off overnight in Microposit Remover 1165. The remaining flake was then etched through via RIE, and the Au hardmask were removed using a solution of potassium monoiodide and iodine (Sigma-Aldrich).

**Chiral Optical Characterization**

The chiral optical characterization was conducted using a custom-built transmission microscope (see SI Note 10). The system was driven by a fiber-coupled supercontinuum white light laser (SuperK FIANIUM from NKT Photonics) set to 5-8% of its maximum power and a repetition rates of 0.7-1.8 MHz. The laser beam was directed through a polarizing beam splitter (PBS), dividing it into horizontal (HP) and vertical (VP) linearly polarized components (2x LPVIS100 from Thorlabs, 550–1500 nm). A quarter wave plate (QWP, RAC4.4.20 from B-Halle, 500–900 nm) was used to generate circularly polarized light (CPL). By blocking the HP or VP path, the polarization could be adjusted between RCP and LCP. This method avoided the need to rotate polarizers or the QWP, which can introduce elliptical polarization if not carefully controlled. The QWP was positioned directly below the objectives to prevent reflections from mirrors, which can convert CPL into elliptically polarized light. The light was condensed onto the sample using an 10x objective (Olympus PLN, NA=0.25) for normal incidence scaling factor measurements and an 20x objective (Olympus PLN, NA = 0.40) for incidence angle measurements. The light was collected using a 60x objective (Nikon MRH08630, NA = 0.7). The beam was condensed to illuminate the entire metasurface area of 30 µm x 30 µm. The incidence angle measurements were performed by rotating the sample holder. For the measurement of co- and crosspolarization terms, a chiral analyzer consisting of a QWP (AQWP05-580 from Thorlabs, 350–850 nm) and a linear polarizer (WP25M-UB from Thorlabs, 250–4000 nm) were installed after the collection objective. A flip mirror was used to direct the light either directly to a CCD camera or to a spectrometer via a multimode fiber (Thorlabs M15L05, core size: 105 µm, NA = 0.22). A spectrometer from Princeton Instruments with a grating period of 300 g/mm, blaze angle of 750 nm, and spectral resolution of 0.13 nm was used. All spectra were recorded with a binning of 6 lines, an exposure duration of 90 ms, and 20 accumulated spectra. All spectra were normalized using a background measurement taken on the same substrate using the matching angle of incidence and co-/cross polarization.

**k-Space Optical Measurements**

The k-space angle-resolved measurements were obtained using a homebuilt optical setup that operates in back-reflection geometry (see SI Note 10). A halogen lamp (Thorlabs SLS201L) served as the excitation light source. The collimated light was directed through a 50:50 beamsplitter to the 60x magnification objective (NA=0.95) and the sample. The same objective was used to collect the reflected light. A Fourier lens was used to image the back focal plane of the objective. With the assistance of a second lens, we created a 4-f system that projected the back focal plane image onto the slit of the spectrometer. To control the polarization of the light, two linear polarizers (LPVISC100-MP2) and one quarter-wave plate (Thorlabs $\lambda/4$, 400-800 nm) were utilized. As the light passed through the first linear polarizer and the quarter-wave plate rotated at 45 degrees, it became circularly polarized. The reflected light then traversed the quarter-wave plate again, converting it back to linearly polarized light. The orientation of the second linear polarizer acted as an analyzer, enabling the performance of angle-resolved measurements with circular polarization information. The image projected onto the spectrometer slit (Princeton Instruments) will be expanded according to the wavelength, aided by a spectrometer grating (150 g/mm, blaze angle of 800 nm). The resulting image is then projected on the CCD sensor.

## Nonlinear Optics Experiments

Harmonic generation experiments used a 140 fs, 80 MHz mode-locked Ti:Sapphire laser (Chameleon Ultra II) pumping an optical parametric oscillator (Chameleon compact OPO) for generating tunable infrared pulses using its idler output (see SI Note 10). The beam passed a Glan-Taylor prism and a broadband infrared half wave plate (HWP) for linear polarization control, as well as a QWP for experiments with circular polarized excitation. A 10x, 0.25 NA objective focused the beam on the sample, while another 10x, 0.25 NA objective collected the third harmonic. A broadband visible QWP was then used for converting circular polarized emission into linear polarized light, and a following HWP and linear polarizer were used to control the collection polarization or handedness. The signal was detected with a spectrometer (Acton SP2300) using a silicon CCD camera (Pixis 100F). For a sketch see SI Note 10.

## Acknowledgements

Funded by the European Union (EIC, OMICSENS, 101129734, ERC, METANEXT, 101078018). Views and opinions expressed are however those of the author(s) only and do not necessarily reflect those of the European Union or the European Research Council Executive Agency. Neither the European Union nor the granting authority can be held responsible for them. This project was also funded by the Deutsche Forschungsgemeinschaft (DFG, German Research Foundation) under grant numbers EXC 2089/1–390776260 (Germany's Excellence Strategy) and TI 1063/1 (Emmy Noether Program), the Bavarian program Solar Energies Go Hybrid (SolTech) and the Center for NanoScience (CeNS). This work was supported by the Australian Research Council (Grant No. DP210101292) and the International Technology Center. Indo-Pacific (ITC IPAC) via Army Research Office (contract FA520923C0023). T.P. acknowledges support from the Bavarian project 'Enabling Quantum Communication and Imaging Applications' (EQAP). S.A.M. additionally acknowledges the Lee-Lucas Chair in Physics. M.V.G. acknowledges the support from the Russian Science Foundation (Project No. 23-42-00091, https://rscf.ru/project/23-42-00091/).

## Author contributions

C.H., A.A.A. and A.T. conceived the idea and planned the research. C.H., T.W., M.H. and J.B. contributed to the sample fabrication. C.H., D.G., L.S. and L.S.M. performed optical measurements. T.P. and L.S.M. performed nonlinear optical experiments. C.H., T.W. and A.A.A. conducted the numerical simulations and data processing. A.A.A., M.V.G. and Y.K. developed the theoretical background. S.A.M., M.V.G., Y.K. and A.T. supervised the project. All authors contributed to the data analysis and to the writing of the paper.

**Conflict of interest**

The authors declare that they have no conflict of interest.

# Supporting Information: Chiral Nonlinear Polaritonics with van der Waals Metasurfaces

Connor Heimig*[1], Alexander A. Antonov*[1], Dmytro Gryb[1], Thomas Possmayer[1], Thomas Weber[1], Michael Hirler[1], Jonas Biechteler[1], Luca Sortino[1], Leonardo de S. Menezes[1,2], Stefan A. Maier[3,4], Maxim V. Gorkunov[5,6], Yuri Kivshar[7], and Andreas Tittl[†1]

[1] *Chair in Hybrid Nanosystems, Nanoinstitute Munich, Faculty of Physics, Ludwig-Maximilians-Universität München, Munich, Germany*
[2] *Departamento de Física, Universidade Federal de Pernambuco, Recife-PE, Brazil*
[3] *School of Physics and Astronomy, Monash University, Clayton, Victoria, Australia*
[4] *Department of Physics, Imperial College London, London, UK*
[5] *Shubnikov Institute of Crystallography, National Research Centre "Kurchatov Institute", Moscow, Russia*
[6] *Theoretical Physics and Quantum Technologies Department, National University of Science and Technology 'MISIS', Moscow, Russia*
[7] *Nonlinear Physics Centre, Research School of Physics, Australian National University, Canberra, Australia*

## Contents

---

*These authors contributed equally to this work.
†Andreas.Tittl@physik.uni-muenchen.de

# 1 Unit Cell Design Parameters

While the period, resonator dimensions, and positioning within the unit cell are all significant parameters for qBIC mode formation, the two pivotal design parameters for achieving maximum chirality are illustrated below in Figure S1. These are the height difference, $\Delta h$, and the opening angle, which should be in-line with the theoretical relation $\alpha \sim k\Delta h$ for the maximal chiral response [1]. The height difference is the core parameter for 3D chirality and is established during fabrication. The influence of increasing $\Delta h$ on maximum chirality is initially pronounced before eventually reaching a plateau (Fig. S1a). This indicates that the necessity for ultra-precise etching of the height difference with nanometer precision is reduced, provided that a benchmark $\Delta h$ is surpassed. Therefore, a tendency towards overetching proved an advantageous approach, as it has only very limited downside for the overall mode formation and chirality of the qBIC resonance, in strong contrast to underetching, resulting in too small $\Delta h$.

The sign of the opening angle $\alpha$ regulates the system's preferred handedness, while simultaneously modulating the chirality and linewidth of the resonance (Fig. S1c). To resolve the upper polariton branch at wavelengths below the exciton, an opening angle $\alpha = 12°$ was deemed optimal for the experiment. This angle was selected to broaden the resonance linewidth, thereby ensuring the qBIC mode would be robust against the higher intrinsic material losses of $WS_2$ in this spectral region (Fig. S1b). All simulations and experiments were conducted using positive opening angles, resulting in left-handed structures throughout this work for the sake of clarity.

This allows the design to be optimized in such a way that the qBIC resonance is maximally chiral and has spectral overlap with the A-exciton in bulk $WS_2$ at room temperature (Fig. S1d). Upon switching to the material model including the exciton, the formation of qBIC-driven chiral self-hybridized exciton-polaritons becomes apparent (Fig. S1e). Accordingly, the emergence of upper and lower polariton branches (UPB, LPB) in left-handed transmittance is observed, whereas the right-handed signal only exhibits the excitonic peak. To illustrate the behavior of our maximally chiral qBIC for both the non-excitonic and the excitonic $WS_2$, we introduce the transmission difference $\Delta T = T_{RR} - T_{LL}$ and examine the spectral dispersion of the pure qBIC signal when shifting the resonance via in-plane scaling factor $S$ of the unit cell (Fig. S1f). In the case of the non-excitonic material, the qBIC exhibits a linear trend. In contrast, when moving to the excitonic $WS_2$ material model, an anticrossing pattern emerges, which is characteristic for strongly coupled systems. Furthermore, since the transmission difference is studied, it is evident that this strong coupling only occurs in the left-handed signal, thereby proving the formation of self-hybridized chiral polaritons in the system.

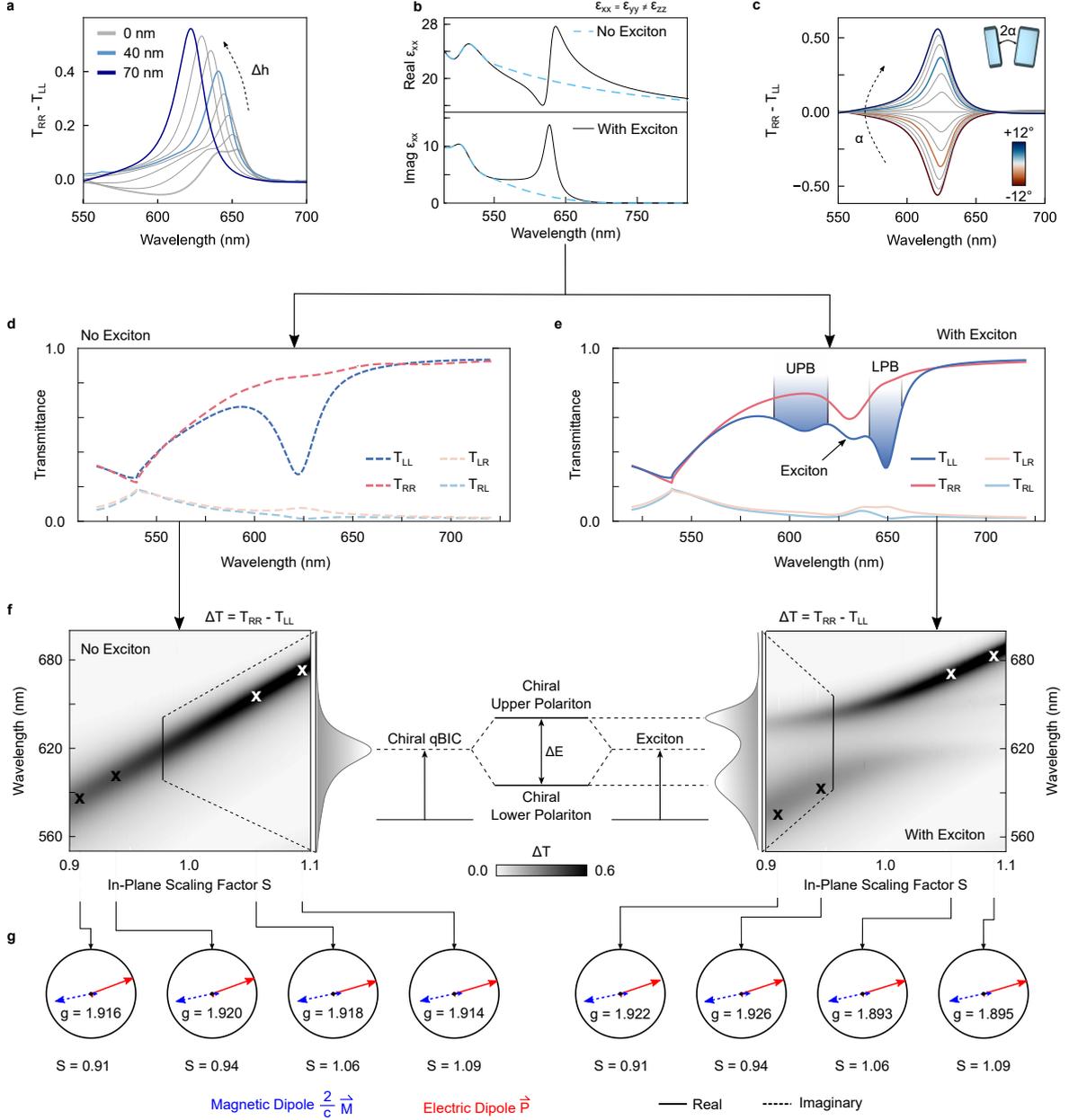

**Figure S1: Unit cell parameters and material model.** Dependence of $T_{RR} - T_{LL}$ on **a** height difference and **c** opening angle $\alpha$. **b** Material Data of $WS_2$ was taken from Mukhbat et al.[2] and adapted to model the material dispersion both with and without the excitonic contribution. **d** Chiral transmittance spectra (LCP/RCP) of the left-handed metasurface without the influence of the exciton for both co- and cross-polarization, highlighting the maximum chirality of the qBIC response. **e** When including the exciton, the chiral transmittance spectra show the emergence of self-hybridized exciton-polaritons. **f** Simulated transmittance differences ($T_{RR} - T_{LL}$) of the chiral $WS_2$ qBIC metasurfaces for different in-plane scaling factors S using the material models with and without the exciton. The simulations with the exciton show a characteristic anticrossing pattern around the exciton. **g** Multipole decomposition of eigenstates with contributions of electric and magnetic dipole moments having almost the same amplitude and phase difference close to $\pi/2$. g is the dissymmetry factor, defined according to Eq. 1 from the main text.

## 2 Chiral Polaritons

The coupling coefficient of an eigenstate of a metasurface with a normally incident plane wave (along the $z$-direction) with a wave number $k = \omega/c$ and polarization unit vector $\mathbf{e}$ can be evaluated as an overlap integral:

$$m_e \propto \int_V \mathbf{J}(\mathbf{r}) \cdot \mathbf{e}\, e^{i\mathbf{k}\cdot\mathbf{r}}\, dV \approx \int_V \mathbf{J}(\mathbf{r}) \cdot \mathbf{e}\, (1 + i\mathbf{k}\cdot\mathbf{r})\, dV$$
$$\approx -i\omega\, \mathbf{P}\cdot\mathbf{e} - i(\mathbf{k}\times\mathbf{M})\cdot\mathbf{e} + \frac{\omega}{6}(e_\alpha)(k_\beta)Q_{\alpha\beta} \tag{1}$$

where $\mathbf{J}(\mathbf{r})$ is an eigenstate displacement current density of the metasurface. The electric and magnetic dipole as well as the electric quadrupole moments are introduced in the standard way[3]:

$$\mathbf{P} = \frac{i}{\omega}\int_V \mathbf{J}(\mathbf{r})\, dV \tag{2}$$

$$\mathbf{M} = \frac{1}{2}\int_V \mathbf{r}\times\mathbf{J}(\mathbf{r})\, dV \tag{3}$$

$$Q_{\alpha\beta} = \frac{3i}{\omega}\int_V \left[ r_\alpha J_\beta(\mathbf{r}) + r_\beta J_\alpha(\mathbf{r}) - \frac{2}{3}\delta_{\alpha\beta}\,\mathbf{r}\cdot\mathbf{J}(\mathbf{r}) \right] dV \tag{4}$$

We consider circularly polarized incident waves with $\mathbf{e}_\pm = \frac{1}{\sqrt{2}}(\mathbf{e}_x \pm i\mathbf{e}_y)$ and reduce Eq. (1) to the following:

$$m_\pm \propto [P_x \pm iP_y] \pm \frac{i}{c}[M_x \pm iM_y] + \frac{i\omega}{6c}[Q_{xz} \pm iQ_{yz}] \tag{5}$$

If the eigenstate current $\mathbf{J}(\mathbf{r})$ flows predominantly in the $\mathbf{xy}$-plane ($J_z \approx 0$), the components of the moments in Eqs. (2–4) contributing to Eq. (5) then approximately reduce to:

$$P_x = \frac{i}{\omega}\int_V J_x(\mathbf{r})\, dV, \quad P_y = \frac{i}{\omega}\int_V J_y(\mathbf{r})\, dV \tag{6}$$

$$M_x = -\frac{1}{2}\int_V z\, J_y(\mathbf{r})\, dV, \quad M_y = \frac{1}{2}\int_V z\, J_x(\mathbf{r})\, dV \tag{7}$$

$$Q_{xz} = \frac{3i}{\omega}\int_V z\, J_x(\mathbf{r})\, dV, \quad Q_{yz} = \frac{3i}{\omega}\int_V z\, J_y(\mathbf{r})\, dV \tag{8}$$

which leads to the following form of the coupling coefficient:

$$m_\pm \propto [P_x \pm iP_y] \pm \frac{2i}{c}[M_x \pm iM_y] \tag{9}$$

A chiral point emitter is a combination of parallel electric $\mathbf{P}$ and magnetic $\mathbf{M}$ point dipoles with a $\pm\pi/2$ phase difference $\mathbf{M} = \pm ic\mathbf{P}$[4]. Conversely, here the extra doubling of $\mathbf{M}$ arising from the nonzero quadrupole moment defines the maximum chirality condition $m_\pm = 0$:

$$\mathbf{M} = \pm \frac{ic}{2}\mathbf{P} \tag{10}$$

We employ the Eigenstate Solver of COMSOL Multiphysics to analyze the polariton fields. We set a constant WS$_2$ permittivity according to its tabular value at a fixed wavelength corresponding to a resonance position in Fig. S1f (with and without excitons, marked as crosses). By calculating the integrals in Eq. (6) and Eq. (7) we analyze the electric dipole moment in the form:

$$\mathbf{P} = \mathbf{P}' + i\mathbf{P}'' \tag{11}$$

where the components of $\mathbf{P}'$ and $\mathbf{P}''$ are purely real. Since the eigenstate is found with an unknown phase $\phi$, we define it as:

$$\tan(2\phi) = \frac{2\mathbf{P}'\mathbf{P}''}{(\mathbf{P}'')^2 - (\mathbf{P}')^2}, \tag{12}$$

so that $\mathrm{Re}(\mathbf{P}e^{i\phi}) \cdot \mathrm{Im}(\mathbf{P}e^{i\phi}) = 0$. Then we plot the real and imaginary parts of $\mathbf{P}e^{i\phi}$ in Fig. S1g and reveal, that the imaginary part of the electric dipole is negligibly small and hardly noticeable. see Fig. S1g). Next we analyze the magnetic dipole:

$$\mathbf{M} = \mathbf{e}_x M_x + \mathbf{e}_y M_y \tag{13}$$

which depends on a coordinate system. Shifting the coordinate origin by $\mathbf{r}' = \mathbf{r}_0 + \mathbf{r}$ where $\mathbf{r}_0 = (x_0, y_0, z_0)^T$ leaves $\mathbf{P}$ unaffected, but changes $\mathbf{M}$. Nevertheless, since the expansion in Eq. (1) was performed in the vicinity of $z = 0$, we can slightly shift the coordinate origin by $\mathbf{r}_0 = (0, 0, z_0)^T$ to match the center of the lower rod (by setting $z_0 = 20\,\mathrm{nm}$). The final magnetic dipole components read as:

$$M_x = -\frac{e^{i\phi}}{2}\left[\int_V z\, J_y(\mathbf{r})\, dV - z_0 \int_V J_y(\mathbf{r})\, dV\right] \tag{14}$$

$$M_y = \frac{e^{i\phi}}{2}\left[\int_V z\, J_x(\mathbf{r})\, dV - z_0 \int_V J_x(\mathbf{r})\, dV\right] \tag{15}$$

Finally, we plot the real and imaginary parts of $2c^{-1}\mathbf{M}e^{i\phi}$ (Fig. S1g) and show that $\mathbf{P}e^{i\phi}$ and $2c^{-1}\mathbf{M}e^{i\phi}$ are collinear and have similar amplitudes. It also becomes evident that the electric dipole contributes primarily through its real part, while the magnetic dipole contributes via its imaginary part, providing the phase difference of $\pi/2$ necessary for a chiral point emitter. Therefore, the above results indicate that $m_- \approx 0$ and the eigenstate is uncoupled from RCP light.

# 3 Three-Dimensional Metasurface Fabrication

In order to achieve a structure featuring resonators with different heights within each metasurface unit cell, a considerable fabrication process is necessary. To this end, $WS_2$ flakes were exfoliated onto fused silica substrates with flake thicknesses ranging from 70 nm to 115 nm. The sample was realigned between the patterning steps done via EBL and RIE (Methods) by using specific gold alignment markers.

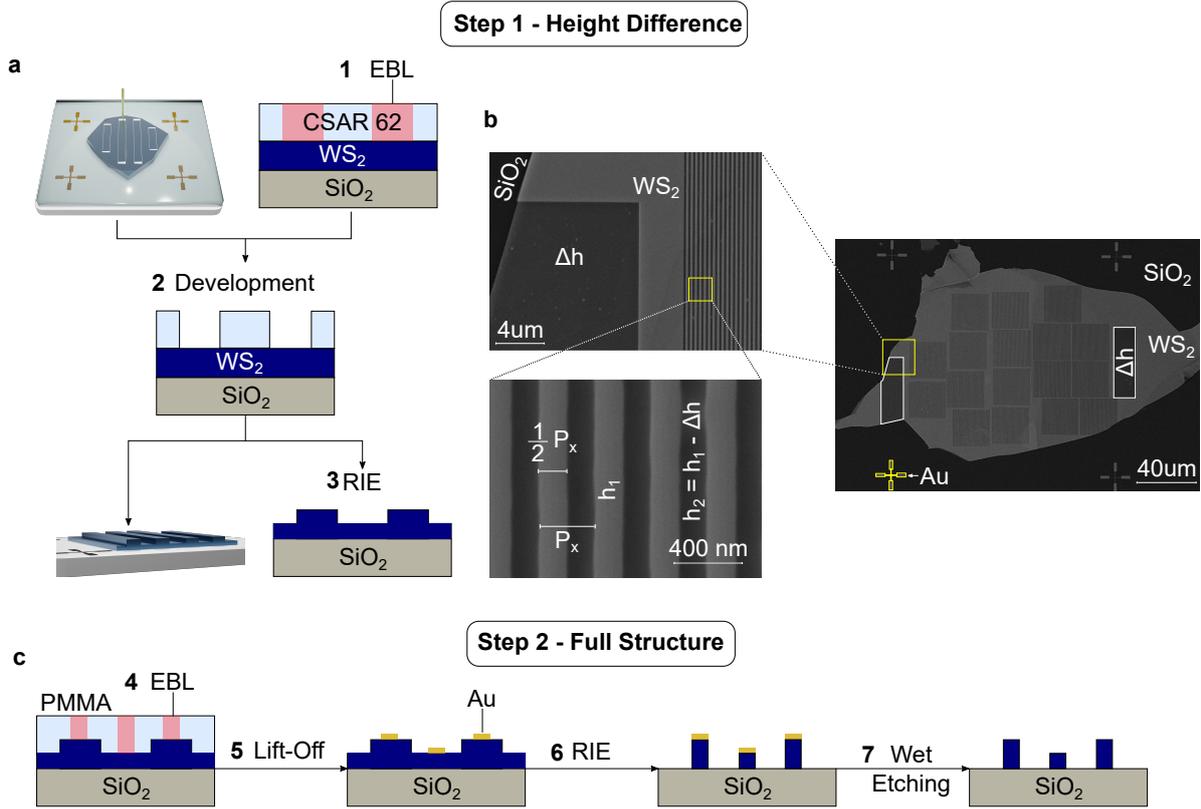

**Figure S2: Experimental realization of 3D chiral $WS_2$ metasurfaces. a** Outline of first inverse EBL step to establish height difference. **b** SEM images of $WS_2$ flakes after first step. The larger rectangles of etched material are used to determine the height difference via profilometer or AFM. **c** Outline of second fabrication step, allowing for simultaneous patterning of both high and low resonator, reducing the needed alignment accuracy to achieve maximum structural fidelity.

To realize the out-of-plane asymmetric metasurfaces presented in this work a two-step fabrication strategy is required that ensures precise alignment between resonators of different heights within a single unit cell. The presented method overcomes these limitations by first introducing a height reference structure directly into the 2D flake (Fig. S2a). In Step 1, a barcode-like pattern corresponding to exactly half of the metasurface unit cell is defined via electron beam lithography (1), developed (2), and the inverse of it transferred into the $WS_2$ flake using reactive ion etching (3). This produces a controlled step height ($\Delta h$) within the flake. As shown in (Fig. S2b), SEM images confirm the lateral pattern fidelity and etch depth, while profilometry verifies the uniformity and reproducibility of $\Delta h$. Larger etched rectangles within the pattern serve as references for measuring the height step. This inverse patterning methodology builds upon previous work.[5]

In Step 2 (Fig. S2c), the full metasurface pattern is defined in a single lithography step (4), overlaid onto the pre-etched barcode such that the high and low resonators fall on the corresponding height regions. In previous approaches, even minimal misalignment between lithography steps could significantly degrade pattern fidelity due to the strict spatial correlation required. [6] In contrast, our design intrinsically tolerates a much broader alignment window: because the barcode spans exactly half the unit cell along x, the full metasurface resonators can be defined without high-precision alignment, while now achieving the highest possible fidelity. Furthermore, since all resonators are written within a single EBL step, perfect vertical and lateral registration between high and low segments is ensured. The only remaining alignment requirement is a lateral (x-axis) overlay of the metasurface pattern onto the barcode. Since the resonators do not span half a period along x (their widths are approximately $\frac{P_x}{4}$ and $\frac{P_x}{3}$ for the slim and wide rods, respectively, and they are rotated by only 12°), any alignment within ±30 nm along the x-axis, which is well within standard EBL tolerances, yields perfeclty accurate 3D chiral metasurface structures. Moreover, because the height differentiation barcode is continuously integrated into the flake along the y-axis, ideal alignment in this direction is intrinsically ensured.

Subsequent processing involves gold deposition and lift-off (5), followed by RIE (6) and wet etching (7) to transfer the pattern into the flake and remove the residual mask. As a result, this technique guarantees perfect intra-unit cell as well as global alignment between resonators and, compared to previous work,[6] offers a simplified process with substantially improved structural fidelity and reproducibility. The strategy effectively eliminates resonator misalignment and enables highly reproducible top-down 3D nanophotonic architectures. Furthermore, since the technique relies exclusively on etching, it is generally applicable to any resonator material system, including exfoliated, chemically grown, or vapor-deposited materials.

# 4 Comparison of Simulation and Experiment

To verify the experimental results from the corresponding section of the main text we compare them with numerical simulations, displaying an excellent agreement of theory and experiment.

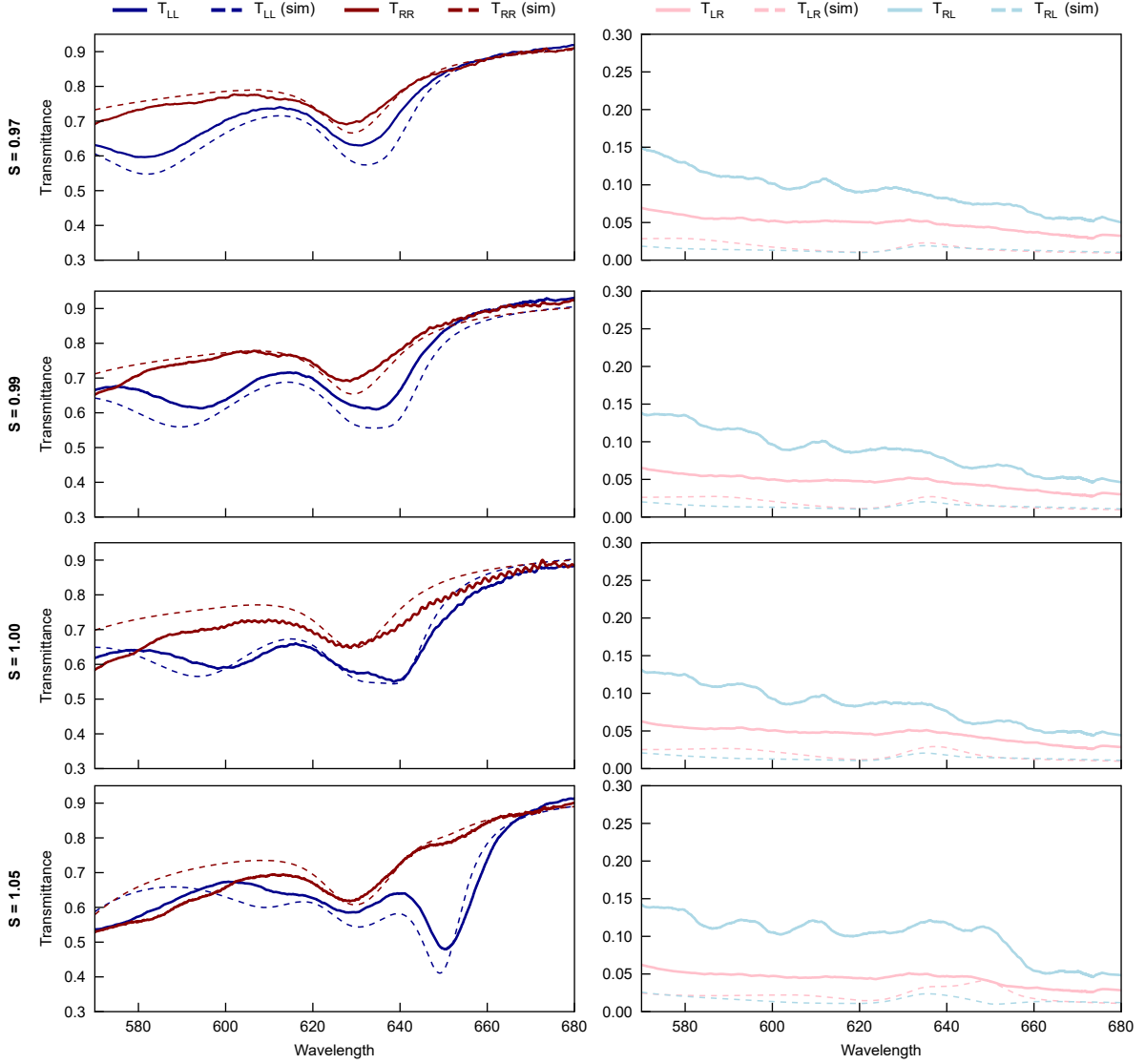

**Figure S3: Chiral transmission under normal incidence.** The unit cell design (for scaling factor S = 1.00) has a periodicity of $P_x = P_y = 370$ nm, two rods with the same length (210 nm) but different widths (130 nm and 63 nm) and heights (30 nm and 80 nm). Each rod is rotated by an angle of $\alpha = 12°$. This design was used for the simulations in Fig. 4 in the main text.

## 5 Temporal Coupled-Mode Theory for BIC–Exciton Coupling

The following section is based on the work of Fan et al.[7] on temporal coupled-mode theory (TCMT) for the Fano resonance in optical resonators. We consider a resonant system coupled to two ports, allowing for both transmission and reflection. The system is excited through port 1, with the incoming wave represented as

$$\mathbf{s}_+ = \begin{pmatrix} s_{1+} \\ 0 \end{pmatrix},$$

and the output waves in port 1 (reflected wave) and port 2 (transmitted wave) represented by

$$\mathbf{s}_- = \begin{pmatrix} s_{1-} \\ s_{2-} \end{pmatrix},$$

where $s_{1+}$, $s_{1-}$, and $s_{2-}$ correspond to the incoming, reflected, and transmitted waves, respectively. In general, the scattering process is described by the scattering matrix $S$ as

$$S(\omega) = C + K\left[i(\omega I - \Omega) + \Gamma\right]^{-1} K^T,$$

where $C$ describes the non-resonant port crosstalk, given by

$$C = e^{i\phi} \begin{pmatrix} r_0 & it_0 \\ it_0 & r_0 \end{pmatrix},$$

with $r_0$ and $t_0$ being the background reflection and transmission coefficients, respectively, constrained by $r_0^2 + t_0^2 = 1$. The phase $\phi$ is a global phase factor. The matrix $\Omega$ contains the complex resonance frequencies $\omega_i = \omega_{0,i} + i\gamma_{\text{int},i}$ on the diagonal, and the near-field coupling rates between the individual modes on the off-diagonal. The matrix $\Gamma$ contains the radiative damping rates and extrinsic mode-coupling rates, defined as $\Gamma_{ij} = \sqrt{\gamma_{\text{rad},i}}\sqrt{\gamma_{\text{rad},j}}$ for $i,j = 1, \ldots, m$, where $m$ is the number of resonant modes. The port-mode coupling is described by the matrix $K$ with elements $K_{nj} = \sqrt{\gamma_{\text{rad},j}}$ for $n = 1, 2$.

The transmission spectrum is obtained from the scattering matrix as

$$T(\omega) = |S_{21}(\omega)|^2,$$

and is used to fit experimental spectra. In our system, as described in [8], the matrices are given by

$$\Omega = \begin{pmatrix} \omega_{\text{BIC}} + i\gamma_{\text{BIC,int}} & \kappa & 0 \\ \kappa & \omega_{\text{Ex}} + i\gamma_{\text{Ex,int}} & 0 \\ 0 & 0 & \omega_{\text{Ex}} + i\gamma_{\text{Ex,int}} \end{pmatrix},$$

$$\Gamma = \begin{pmatrix} \gamma_{\text{BIC,rad}} & 0 & \sqrt{\gamma_{\text{BIC,rad}}\gamma_{\text{Ex,rad}}} \\ 0 & 0 & 0 \\ \sqrt{\gamma_{\text{BIC,rad}}\gamma_{\text{Ex,rad}}} & 0 & \gamma_{\text{Ex,rad}} \end{pmatrix},$$

$$K = \begin{pmatrix} \sqrt{\gamma_{\text{BIC,rad}}} & 0 & \sqrt{\gamma_{\text{Ex,rad}}} \\ \sqrt{\gamma_{\text{BIC,rad}}} & 0 & \sqrt{\gamma_{\text{Ex,rad}}} \end{pmatrix},$$

$$C = e^{i\phi} \begin{pmatrix} r_0 & it_0 \\ it_0 & r_0 \end{pmatrix},$$

where $\kappa$ denotes the coupling strength between the BIC and the exciton.

To accurately fit the coupling strength, we employ an analysis adapted from Nan et al.[9], where we simultaneously fit spectra corresponding to different metasurface scaling factors with shared parameters (Fig. S4a). The exciton position is fixed at 476.6 THz and its linewidth at 8.7 THz, as reported in [8]. The uncoupled BIC dispersion is approximated as a linear function, with its slope and offset treated as shared fit parameters. Due to increased material absorption at smaller scaling factors, the BIC linewidth increases, which is also modeled as a linear function of the scaling factor (Fig. S4b). We evaluate the coupling strength $\kappa$ at the scaling factor $S \approx 1.035$, where the BIC and exciton resonances overlap, and calculate the generalized Rabi frequency as

$$\Omega_R = 2\sqrt{\kappa^2 - \frac{(\gamma_{\text{BIC}} - \gamma_{\text{Ex}})^2}{4}}.$$

The polariton branches are described by

$$\omega_\pm = \frac{\omega_{\text{BIC}} + \omega_{\text{Ex}}}{2} + i\frac{\gamma_{\text{BIC}} + \gamma_{\text{Ex}}}{2} \pm \sqrt{\kappa^2 - \frac{1}{4}\left(\gamma_{\text{BIC}} - \gamma_{\text{Ex}} + i(\omega_{\text{BIC}} - \omega_{\text{Ex}})\right)^2},$$

which show excellent agreement with our experimental transmittance spectra (Fig. S4c).

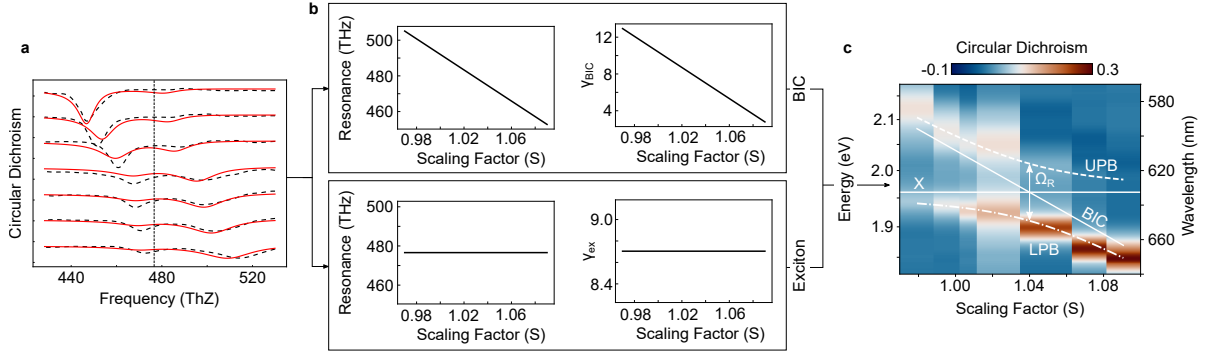

**Figure S4: Temporal coupled-mode theory for BIC–exciton coupling. a** Simultaneous TCMT fit of metasurfaces with multiple scaling factors. **b** Scaling factor dependence of resonance position and damping rate. The BIC dispersion is linear for both the resonance position and damping rate (upper panel), while the exciton position and linewidth remain constant (lower panel). **c** Branches of strongly coupled exciton-polaritons. The colormap shows the circular dichroism from experimental transmittance spectra, where the overlayed polariton branches are calculated from above fit parameters, showing excellent agreement, corroborating the accuracy of our fit approach.

# 6 Strong Coupling Fits in k-Space

To further corroborate the validity of our TCMT fits, we conduct additional fits of experimental reflectance spectra for different scaling factors in the k-space. We estimate the upper limit of the Rabi splitting $\Omega_\mathrm{R}$ by using a simplified Hamiltonian

$$\Omega = \begin{pmatrix} \omega_\mathrm{BIC} & \kappa \\ \kappa & \omega_\mathrm{Ex} \end{pmatrix},$$

which leads to the following simplified polariton dispersions

$$\omega_\pm(k) = \frac{\omega_\mathrm{BIC}(k) + \omega_\mathrm{Ex}}{2} \pm \sqrt{\kappa^2 + \frac{1}{4}\left(\omega_\mathrm{BIC}(k) - \omega_\mathrm{Ex}\right)^2}.$$

While in simulation, the dispersion of the qBIC $\omega_\mathrm{BIC}(k)$ is known and used to fit the polariton branches, in experiment, we approximate the dispersion with the quadratic relation

$$\omega_\mathrm{BIC}(k) = Ak^2 + C(S),$$

with A and C being fit parameters. We further assume that the parameter $A$ is the same for all scale factors probed and that $C$ follows a linear relationship with the scaling factor $S$: $C(S) = C^*S + C^{**}$. Similarly to the above fit of multiple spectra with shared fit parameters, we share the parameter $A$ as well as the slope and offset of the parameter C over a scaling factor range from $S = 0.95$ to $S = 1.11$. Fixing the coupling strength $\kappa$ to $13.07\,\mathrm{THz}$, taken from the above TCMT fits, yields excellent agreement of the polariton fits with the strongly coupled experimental data (Fig. S5).

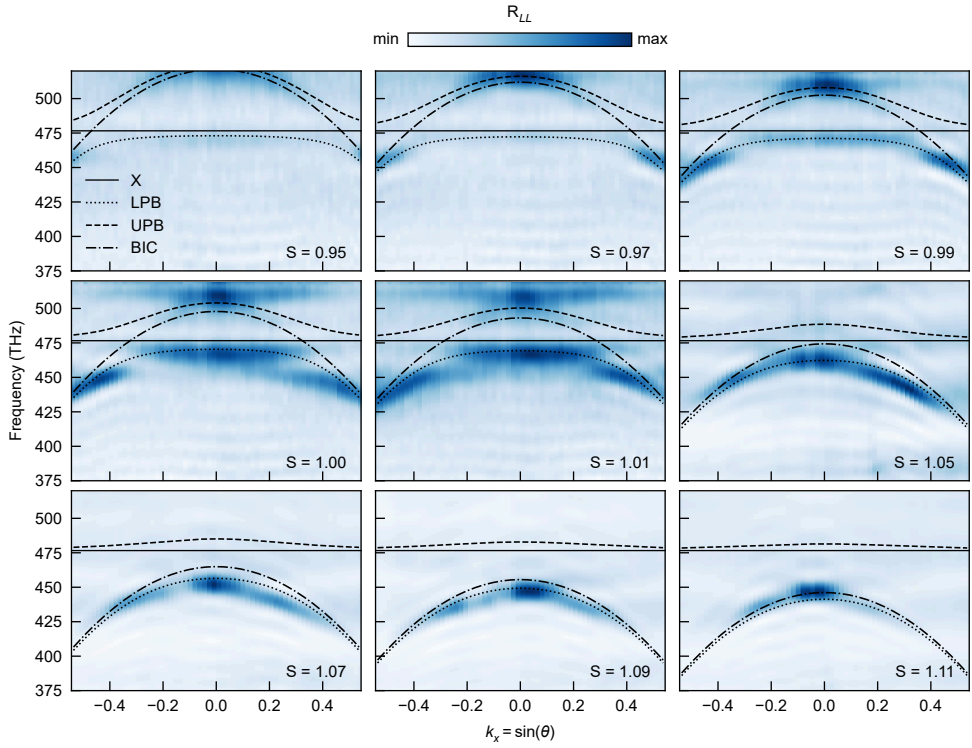

**Figure S5: Experimental k-space polariton fits.** Upper- and lower polariton dispersions fitted to experimental angle-resolved reflectance spectra for left-circularly polarized light ($R_{LL}$) with a fixed coupling strength $\kappa$ and parabolic dispersion of the quasi-BIC mode, which is linearly shifted with the scaling factor $S$.

# 7 THG from Unpatterned 105 nm Thick WS$_2$ Flake

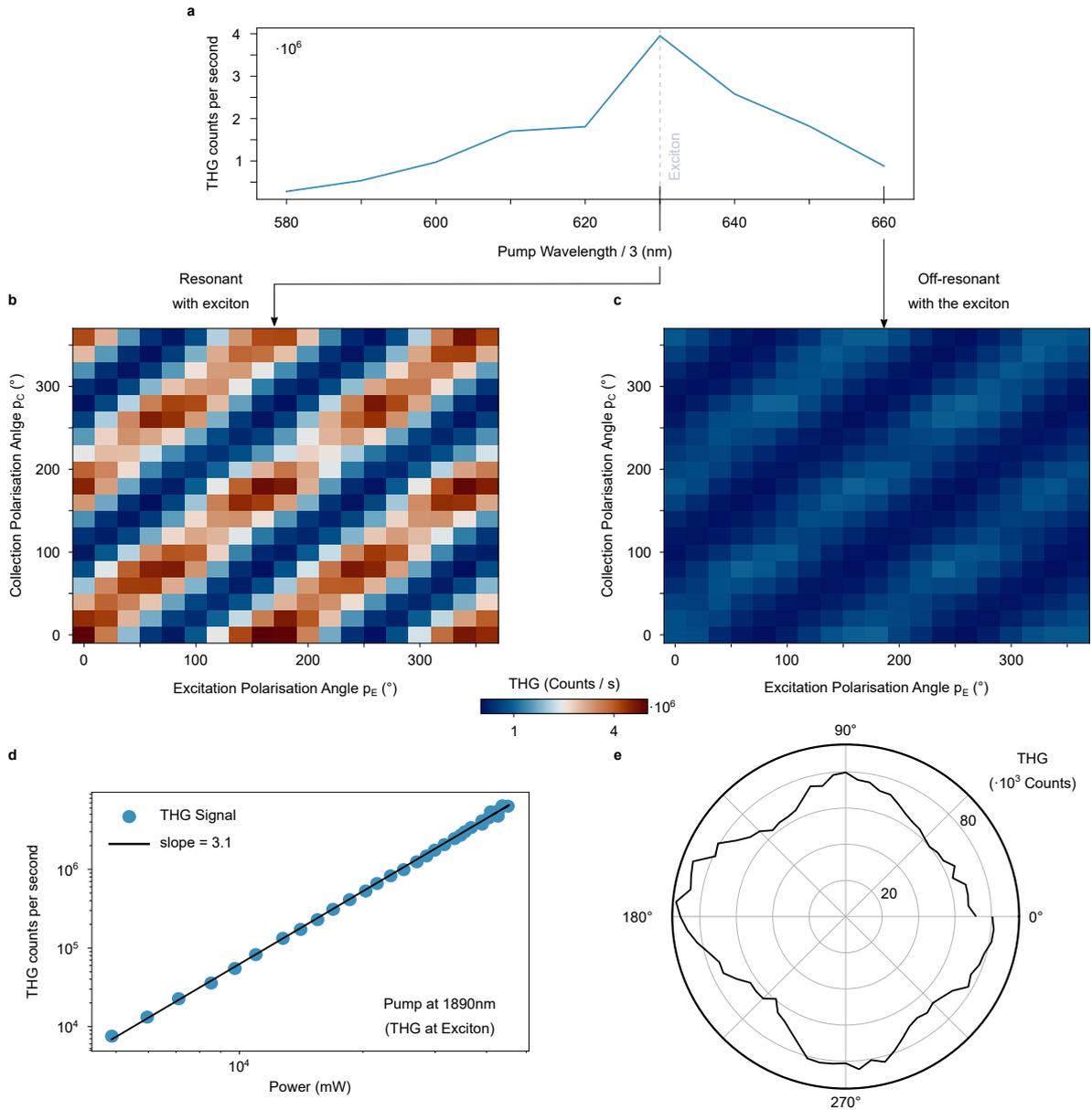

**Figure S6: THG of unpatterned 105 nm thick WS$_2$ Flake excited with linearly polarized light. a** THG counts per second vs. pump wavelength/3, showing intrinsic enhancement of nonlinear generation near the exciton. **b** Polarization dependence of resonant THG (pump: 1890 nm, 15 mW) reveals no preferred polarization in unpatterned material. **c** Polarization dependence of off-resonant THG (pump: 1980 nm, 15 mW) similarly shows no polarization preference, with reduced THG confirming exciton-enhanced generation. **d** Power dependence of resonant THG from unpatterned WS$_2$ shows a slope of 3.1 on a double-log scale. **e** Normalized polarization-dependent THG from unpatterned WS$_2$ at 1860 nm, 15 mW pump.

# 8 THG from WS$_2$ Metasurfaces Excited with Linearly Polarized Light

We characterized the power dependence of the nonlinear signal to verify the order of the generated harmonic. To this end, we plotted the signal intensity as a function of pump power on a double-logarithmic scale (Fig. S7). The slope of the linear fit to this data determines the harmonic order. The extracted slope closely matches 3, confirming that the signal arises from a third-order process (in the present case THG). Next, we examined the influence of the qBIC metasurface by sweeping the excitation polarization and comparing the THG response from both patterned and unpatterned WS$_2$ (Fig. S7b). In the x-y plane, the unpatterned WS$_2$ is isotropic (with anisotropy only along the z-axis), resulting in no preferred polarization for the THG signal. The polarization response of the unpatterned sample therefore forms a circle. In contrast, the polarization-dependent THG from the qBIC metasurface exhibits a pronounced dumbbell shape, indicating a strong polarization dependence. The maximum THG signal aligns with the preferred excitation polarization of the qBIC mode in the double-rod unit cell (parallel to the principal axes), confirming that the THG is mediated by the qBIC. When the excitation polarization is orthogonal to the BIC mode - where the qBIC resonance vanishes - the THG signal nearly disappears, providing further evidence that the nonlinear response is governed by the qBIC. For further confirmation and to get a more comprehensive picture, the THG is swept for the entire parameter space spanned by the excitation and collection polarizations (p$_E$ and p$_C$, respectively) (Fig. S7d). The data again indicates that there is clear preferred excitation polarization that maximizes THG, which in turn is consistent with the polarization that leads to the strongest qBIC formation.

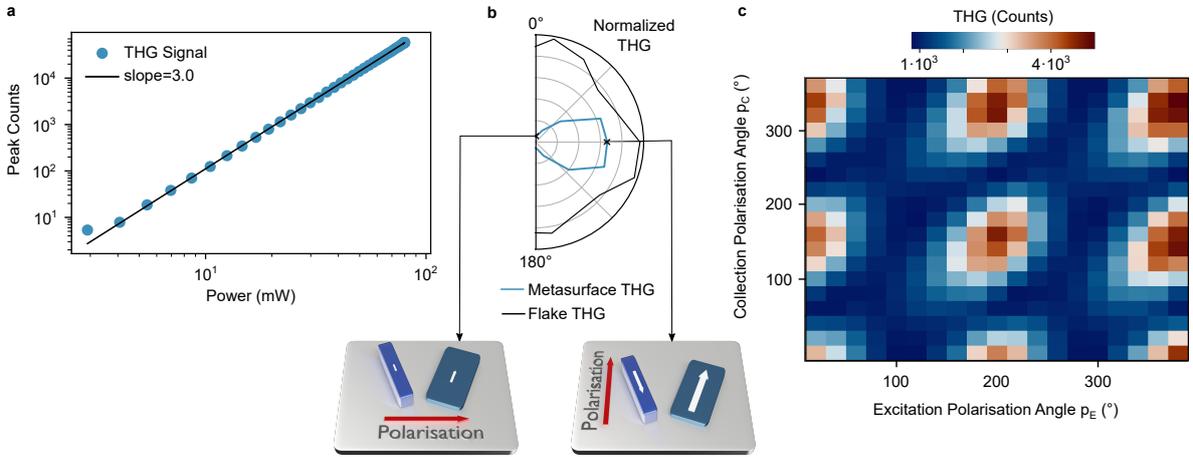

**Figure S7: THG under linearly polarized illumination. a** The power dependence of the higher harmonic signal from the chiral WS$_2$ metasurfaces shows a linear trend with a slope of 3 when plotted on a double log-scale, typical for third harmonic signal. **b** Normalized polarization dependent THG signal from the WS$_2$ metasurface excited with linearly polarized light. The metasurface-enhanced THG signal shows a characteristic dumbbell shape that is absent from the unpatterned WS$_2$, highlighting the importance of the qBIC resonance for THG generation in the structure. The respective lines do not close due to the beam shift caused by changing the polarization. **c** Influence of excitation and collection polarization on the THG signal.

# 9 Polarization Analysis of THG from WS$_2$ Metasurfaces

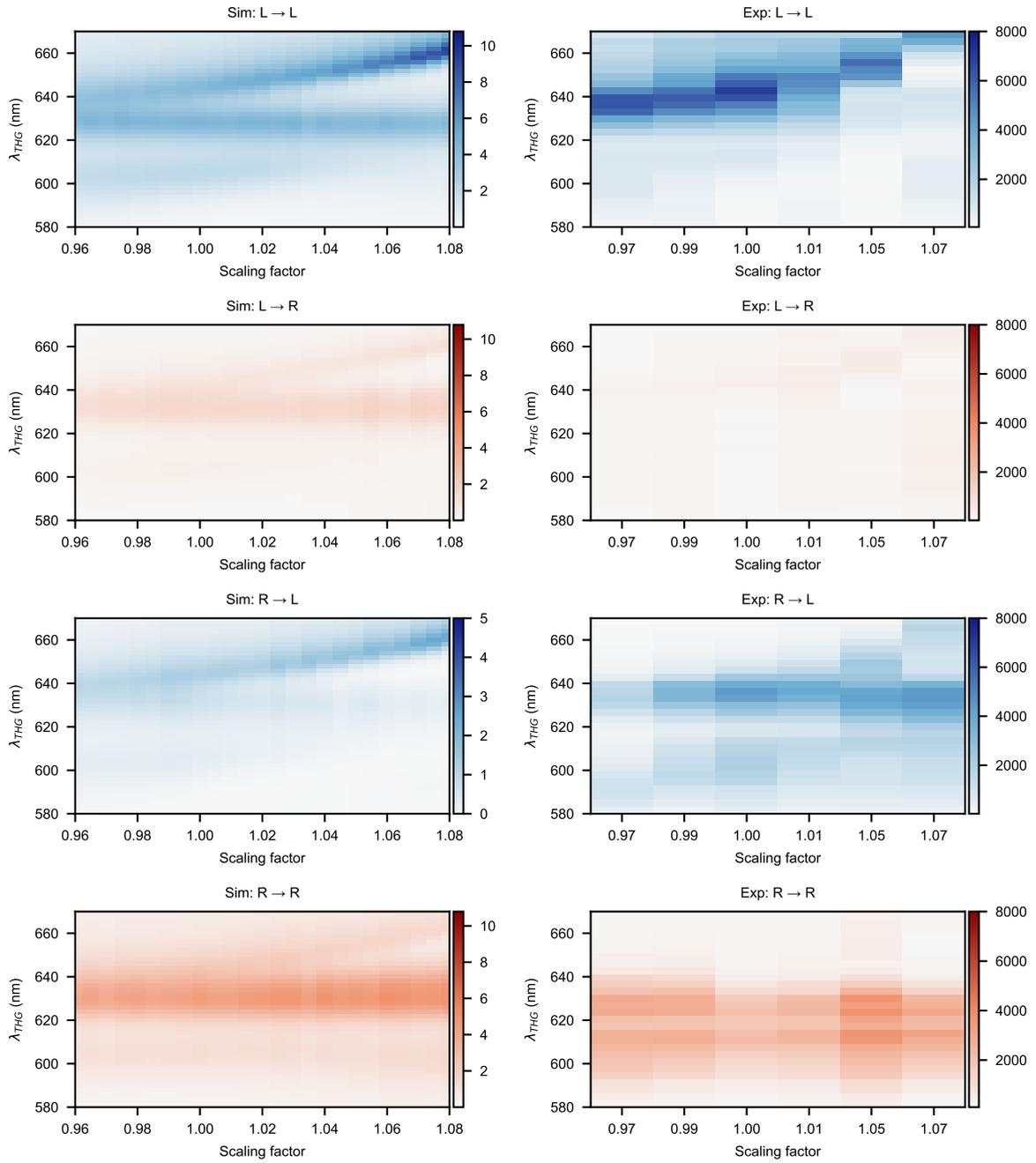

**Figure S8: THG chiral pump polarization** THG spectra in counts per seconds (experiment) and arbitrary units (simulations) for different polarizations of excitation.

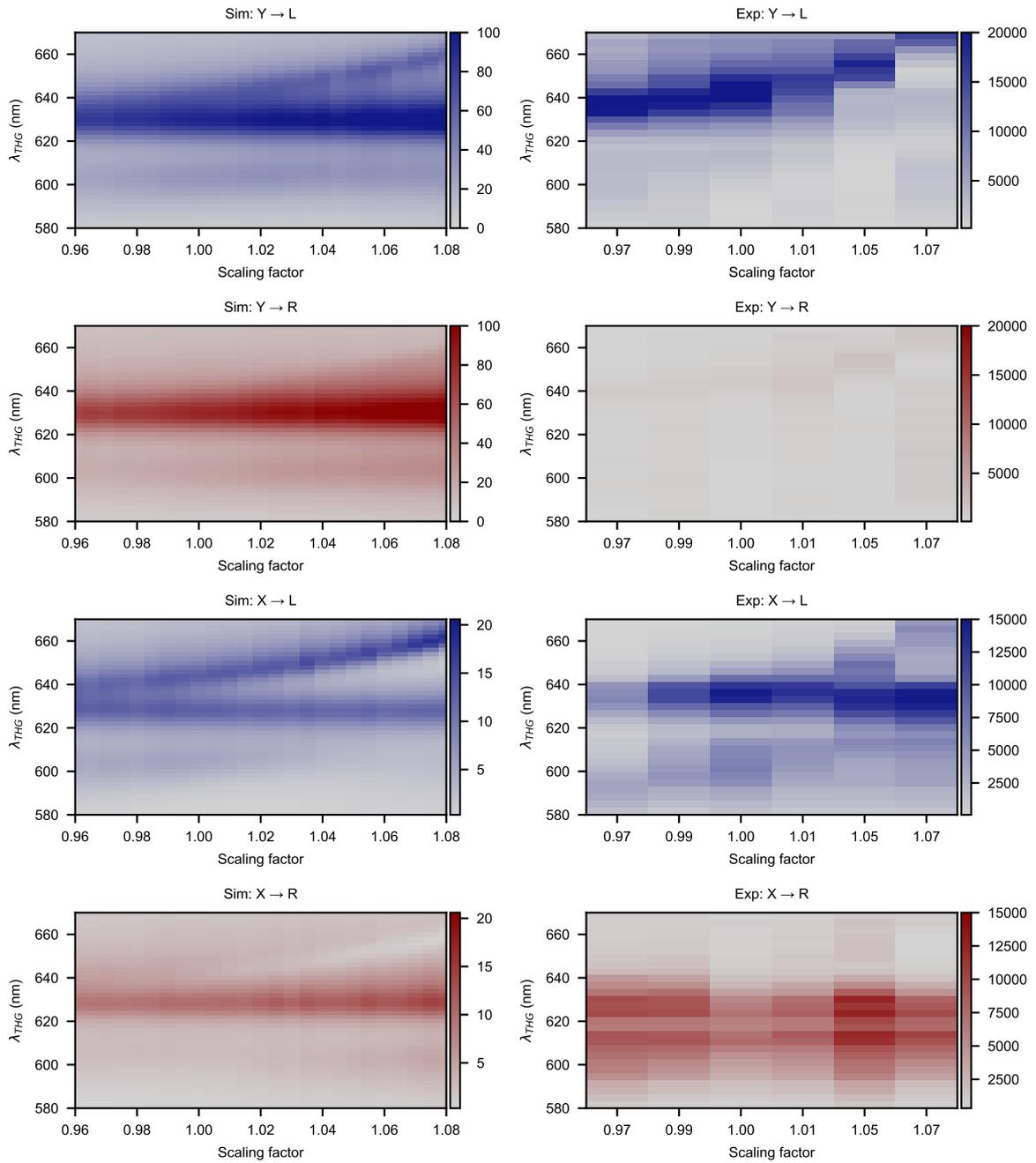

**Figure S9: THG linear pump polarization** THG spectra in counts per seconds (experiment) and arbitrary units (simulations) for different polarizations of excitation.

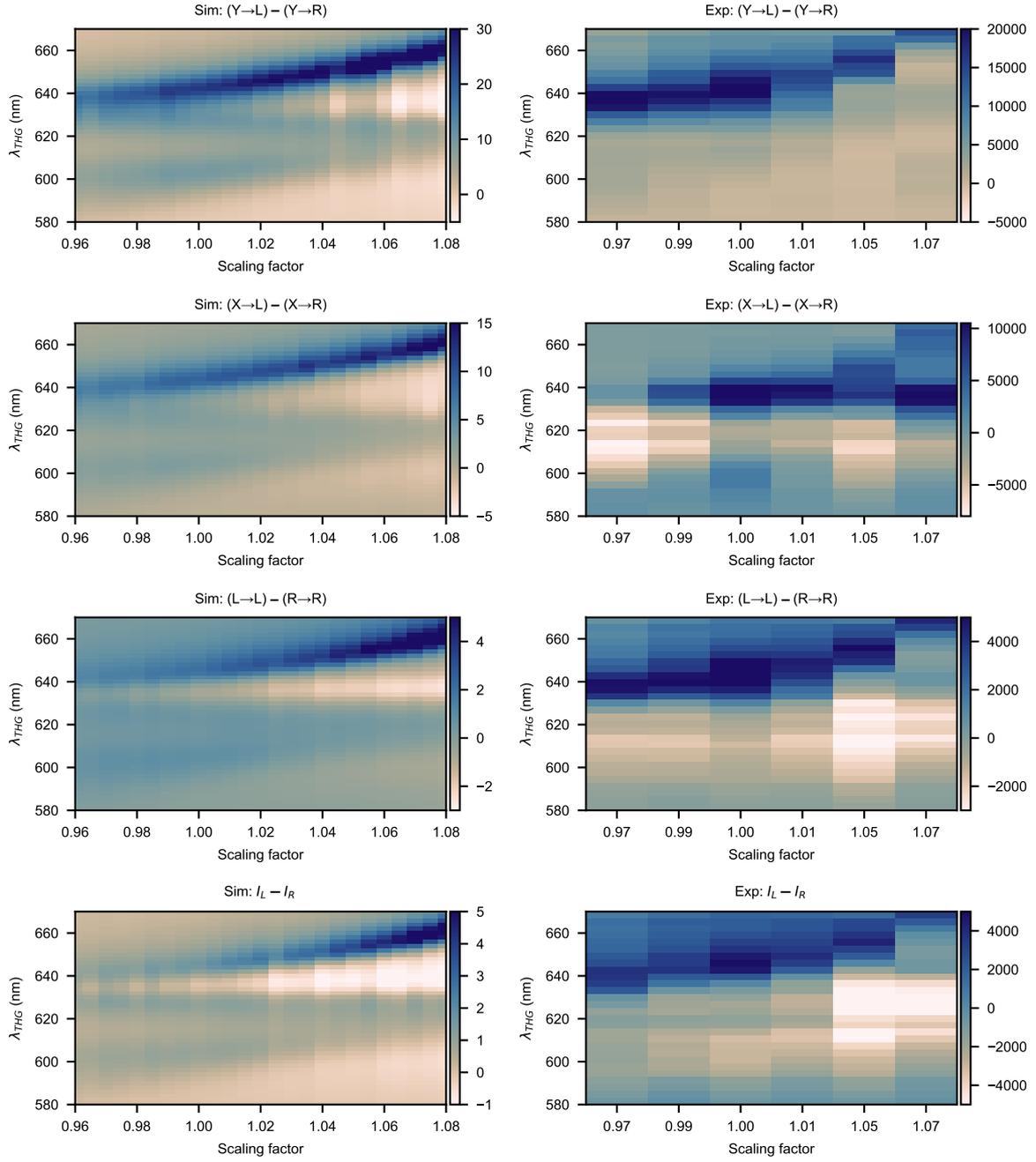

**Figure S10: THG intensity differences** THG spectra in counts per seconds (experiment) and arbitrary units (simulations) for different polarizations of excitation. $I_L$ ($I_R$) is the total THG intensity collected for LCP (RCP) pumping: $I_L$ = L to L + L to R ($I_R$ = R to R + R to L).

# 10 Sketches of Measurement Setups

**Details on Chiral Transmission Measurements**

A sketch of the chiral measurement setup is shown below. The excitation objective was switched to 20x magnification for the measurements at oblique angle of incidence to improve possible alignment quality.

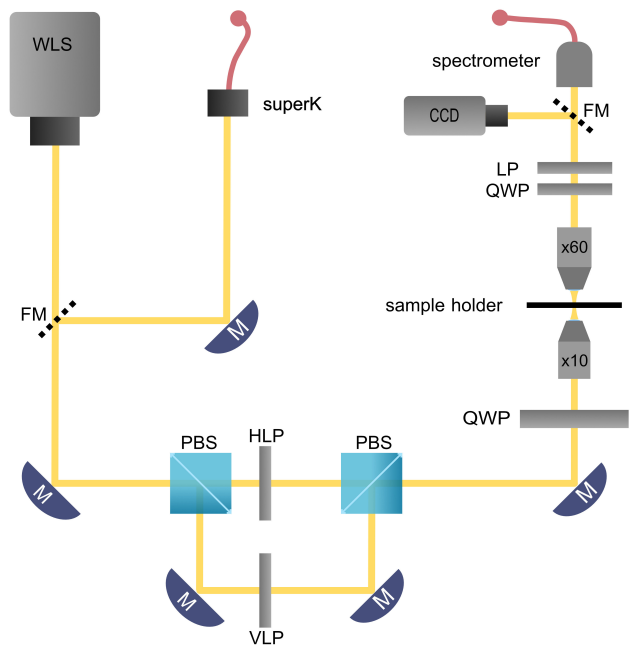

**Figure S11: Sketch of chiral measurement setup.** Used abbreviations are WLS: white light source, M: mirror, FM: flip mirror, PBS: polarizing beam splitter, HLP (VLP): horizontal (vertical) linear polarizer, QWP: quarter-wave plate, LP: linear polarizer

## Details on Back Focal Plane spectroscopy Measurements

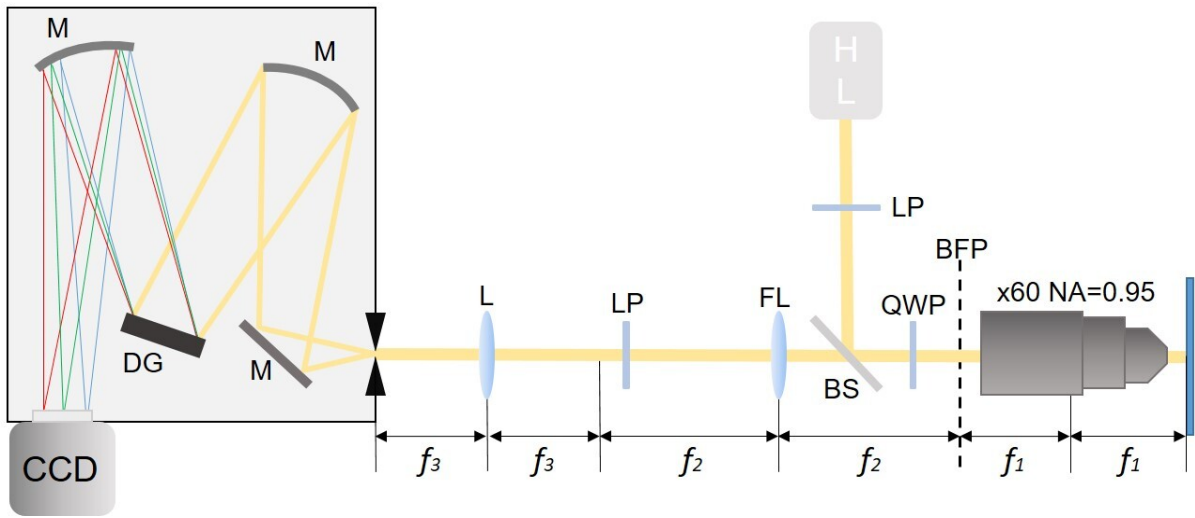

**Figure S12: Sketch of back focal plane spectroscopy setup.** Schematic of the k-space angle-resolved reflection setup. A halogen lamp (HL) provides collimated white light, which is circularly polarized via a linear polarizer (LP) and quarter-wave plate (QWP) before being directed to the sample through a 60×, 0.95 NA objective via a 50:50 beamsplitter (BS). Reflected light, collected by the same objective, passes back through the QWP and an analyzer LP to resolve polarization states. A Fourier lens (FL) images the back focal plane (BFP), and a 4-f system relays the image onto the entrance slit of a spectrometer. The spectrally dispersed signal (M: mirror; DG: diffraction grating) is then recorded by a CCD camera.

## Details on Non-Linear Measurements

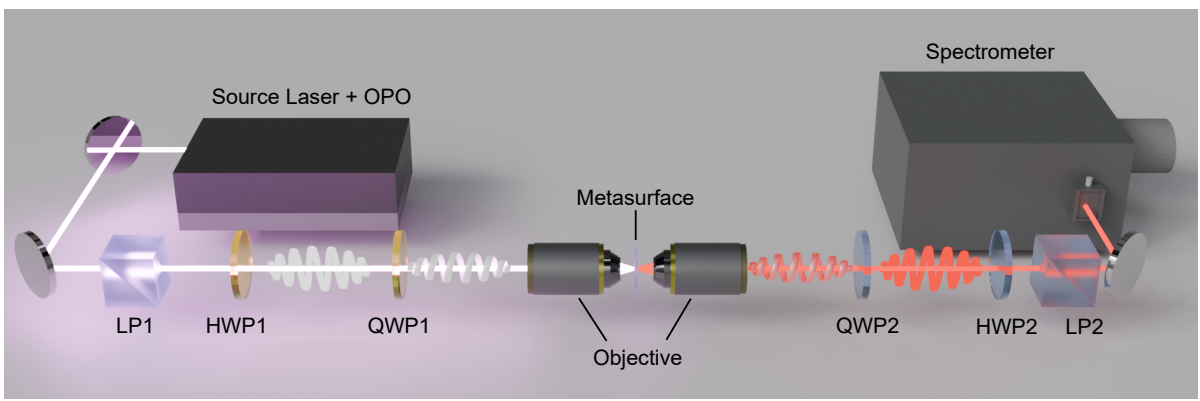

**Figure S13: Sketch of chiral nonlinear measurements setup.** For linear THG measurements, the QWPs are removed. LP1 and LP2: Glan-Taylor-Prisms, HWP1 and QWP1: waveplates for the infrared, QWP2 and HWP2: visible waveplates.

# References

1. Gorkunov, M. V., Antonov, A. A., Tuz, V. R., Kupriianov, A. S. & Kivshar, Y. S. Bound states in the continuum underpin near-lossless maximum chirality in dielectric metasurfaces. *Advanced Optical Materials* **9,** 2100797 (2021).

2. Munkhbat, B., Wróbel, P., Antosiewicz, T. J. & Shegai, T. O. Optical constants of several multilayer transition metal dichalcogenides measured by spectroscopic ellipsometry in the 300–1700 nm range: high index, anisotropy, and hyperbolicity. *ACS Photonics* **9,** 2398–2407 (2022).

3. Evlyukhin, A. B., Fischer, T., Reinhardt, C. & Chichkov, B. N. Optical theorem and multipole scattering of light by arbitrarily shaped nanoparticles. *Physical Review B* **94,** 205434 (2016).

4. Baranov, D. G., Schäfer, C. & Gorkunov, M. V. Toward molecular chiral polaritons. *ACS Photonics* **10,** 2440–2455 (2023).

5. Biechteler, J. *et al.* Fabrication Optimization of van der Waals Metasurfaces: Inverse Patterning Boosts Resonance Quality Factor. *Advanced Optical Materials,* 2500920 (2025).

6. Kühner, L. *et al.* Unlocking the out-of-plane dimension for photonic bound states in the continuum to achieve maximum optical chirality. *Light: Science & Applications* **12,** 250 (2023).

7. Fan, S., Suh, W. & Joannopoulos, J. D. Temporal coupled-mode theory for the Fano resonance in optical resonators. *Journal of the Optical Society of America A* **20,** 569–572 (2003).

8. Weber, T. *et al.* Intrinsic strong light-matter coupling with self-hybridized bound states in the continuum in van der Waals metasurfaces. *Nature Materials* **22,** 970–976 (2023).

9. Nan, L. *et al.* Angular dispersion suppression in deeply subwavelength phonon polariton bound states in the continuum metasurfaces. *Nature Photonics,* 1–9 (2025).